\documentstyle[12pt,aasms4]{article}

\lefthead{Sil'chenko et al.}
\righthead{Chemicallly Decoupled Nuclei in NGC 4216 and NGC 4501}

\begin{document}

\title{Chemically Decoupled Nuclei in the Spiral Galaxies NGC~4216
and NGC~4501}

\author{O. K. Sil'chenko}
\affil{Sternberg Astronomical Institute, Moscow, 119899 Russia\\
       Isaac Newton Institute, Chile, Moscow Branch\\
     Electronic mail: olga@sai.msu.su}

\and

\author{A. N. Burenkov and V. V. Vlasyuk}
\affil{Special Astrophysical Observatory, Nizhnij Arkhyz,
    357147 Russia\\
    Electronic mail: ban@sao.ru, vvlas@sao.ru}

\begin{abstract}
By using bidimensional spectral data obtained at the 6m telescope
for the Virgo spirals NGC~4216 and NGC~4501, we have found chemically
distinct metal-rich nuclei in these galaxies. Under the assumption
of equal ages for the nuclear and bulge stellar populations, the
metallicity difference between the nuclei and their environments in
the galaxies is estimated as a factor of 2. But we have also found
an age difference between the nucleus and the bulge in NGC~4216:
age-metallicity disentangling on the diagrams (H$\beta$, Mgb),
(H$\beta$, [MgFe]),
and (H$\beta$, $<\mbox{Fe}>$) results in an age estimate for the
nucleus of 8--12 billion years, the bulge being older by a factor
of 1.5--2; and the self-consistent metallicity difference estimate
is then a factor of 3. The solar magnesium-to-iron ratios in the
galactic nuclei show evidence for a long duration of the secondary
nuclear star formation bursts which produced the chemically distinct
stellar subsystems. Detailed morphological and kinematical analyses
made for the stellar and gaseous structures in the centers of
NGC~4216 and 4501 have revealed the presence of circumnuclear
stellar-gaseous disks with radius of some hundreds parsecs
which demonstrate fast axisymmetric rotation
and lie exactly in the planes of the main galactic disks.

\end{abstract}

\keywords{galaxies: spiral --- galaxies: individual (NGC 4216) ---
galaxies: individual (NGC 4501) --- galaxies: abundances --- galaxies:
nuclei --- galaxies: kinematics and dynamics --- galaxies: structure}

\section{Introduction}

We search for chemically distinct stellar nuclei in disk
galaxies for several years, and this search has been successful:
chemically distinct nuclei are found in a dozen of disk galaxies
(\cite{sil97}). From the beginning we tried to find a relation
between the presence of chemically distinct nuclei and the galactic
structure. At first we have noticed that isophote twist is absent,
as a rule, in the galaxies with chemically distinct nuclei
(\cite{we92}); so we have associated the chemically distinct nuclei
with central mass concentrations within axisymmetric bulges and have
thought them to be the result of the internal evolution. But later some
arguments have been accumulated that the presence of chemically
decoupled nucleus is often accompanied by kinematically decoupled
gaseous subsystems, such as polar rings, so it may be a consequence
of a past interaction with some gas accretion and subsequent secondary
star formation burst in the nucleus (\cite{silvb97,sil98}).

The chemically distinct nuclei in elliptical galaxies are more widely
known, and their presence is often related to nuclear compact stellar
disks within main galactic bodies. Bender and Surma (\cite{bs92})
reported the discovery of a magnesium index enhancement in dynamically
decoupled cores of four elliptical galaxies. Since they deal with
dynamically hot elliptical galaxies, the chemically decoupled cores
have been easily identified as being disks from dynamical arguments,
namely, from their fast rotation and, in two cases, from a slight
decrease of stellar velocity dispersion toward the centers. Later for
two galaxies a photometric study has confirmed the presence of the
nuclear stellar disks with appropriate sizes (NGC~5322: \cite{scb95};
NGC~4365: \cite{sb95}). Recently another two elliptical galaxies,
NGC~4816 and IC~4051 in the Coma cluster, were found to possess
chemically (and kinematically) distinct cores which are in fact
central stellar disks (\cite{mehetal}). The hypothesis that the
kinematically and chemically distinct cores in elliptical galaxies
are nuclear stellar
disks formed via dissipational major merger events has become
popular. However one must admit that there exists alternative points of
view: Carollo et al. (\cite{car97}) who used the HST WPFC2 data have not
found any nuclear disks in the elliptical galaxies with kinematically
and chemically distinct cores NGC~2434 and NGC~7192 and have
concluded that magnesium-rich cores "may arise from a large variety
of morphological and dynamical structures".

Spiral galaxies were studied less thoroughly, but we know one
example which was considered in detail. It is NGC~4594, which has been
investigated both spectroscopically and photometrically more than
once. Particularly, results of bidimensional spectroscopy presented by
Emsellem et al. (\cite{ems96}) are of great interest for us in the frame
of our topic. NGC~4594 has an inner disk with the radius of 11\arcsec\
which is disconnected to the more outer main disk of the galaxy --
this was already shown by Burkhead (\cite{burk}). Two-dimensional maps of
absorption-line indices Mgb, Fe5270, Fe5335, and Fe5406 obtained by
Emsellem et al. (\cite{ems96}) have revealed the presence of a chemically
distinct nucleus in NGC~4594; but, interestingly, the magnesium-index
difference between the nucleus and the inner disk is smaller by a factor
of 2 than the corresponding difference between the nucleus and the bulge
(along the minor axis), and there are no difference of iron indices
between the nucleus and the inner disk. In some sense, just
the inner disk in NGC~4594 represents a chemically distinct core.

So, we suspect that the presence of chemically distinct nuclei in
spiral galaxies may be related to their inner stellar disks; may be,
it is the same thing. But the sample of spiral galaxies with
decoupled circumnuclear stellar disks is not larger (perhaps, even
smaller) than the sample of spiral galaxies with chemically
distinct nuclei. An alternative starting point is proposed by a recent
work of Rubin et al. (\cite{rubin97}) who have reported the discovery
of 14 rapidly rotating, discrete circumnuclear {\it gaseous} disks
in Virgo cluster spiral galaxies. Among those, NGC~4216 possesses
one of the most extended solid-body rotating circumnuclear gaseous
disks: Rubin et al. (\cite{rubin97}) give its radius equal
to 10\arcsec\ (or 800 pc) and its angular rotation velocity equal
to 28 km/s/arcsec, or 350 km/s/kpc.
Since NGC~4216 is also included into our list of candidates for
possessing a chemically distinct nucleus (\cite{sil94}), we think
it to be an appropriate object which may contain a distinct nuclear
{\it stellar} disk; due to the large extension of the circumnuclear
gaseous disk, the stellar disk may be also large enough and so may
be resolved
under moderate seeing conditions. Another Virgo spiral, NGC~4501, is
not mentioned by Rubin et al. (\cite{rubin97}) as having a circumnuclear
gaseous disk; moreover, Keel (\cite{keel}) stated that its central
emission is confined to the nucleus and does not show any disk
structure.
But it is also in our list of candidates for possessing a chemically
distinct nucleus (\cite{sil94}), and we have decided to study it
more carefully together with NGC~4216. We have undertaken bidimensional
spectroscopy of the central parts of these galaxies. Section~2 describes
the details of our spectral observations and of other observational
data involved into our analysis. In Section~3 we discuss the morphology
of surface brightness distributions in
NGC~4216 and NGC~4501. Section~4 deals with the kinematics of the gas
and stars in the central parts of these galaxies.
Section~5 presents the discovery of their chemically
distinct nuclei. Finally, Section~6 contains a brief summary and
conclusions.

\section{Observations and Data Reductions}

We have undertaken spectral investigation of NGC~4216 and NGC~4501 at
the 6m telescope of the Special Astrophysical Observatory of the
Russian Academy of Sciences. A full journal of the spectral
observations is presented in Table~1.

\begin{table}
\caption[ ] {Spectral observations of NGC~4216 and 4501}
\begin{flushleft}
\begin{tabular}{lcllllcl}
\tableline
Date & Galaxy & Configuration & Exposure & Field & Sp. range
& PA & Seeing \\
\tableline
12.05.96 & NGC~4501 & MPFS+CCD $520\times580$ & 45 min &
$10\arcsec\times16\arcsec$ & 4800--5400~\AA\ & $ 1^\circ$ &
2\arcsec \\
12.05.96 & NGC~4501 & MPFS+CCD $520\times580$ & 40 min &
$10\arcsec\times16\arcsec$ & 6200--6900~\AA\ & $ 1^\circ$ &
2\arcsec \\
15.07.96 & NGC~4501 & LS+CCD $520\times580$ & 30 min &
$2\arcsec\times2.5\arcmin$ & 6200--7000~\AA\ & $137^\circ$ &
1.3\arcsec \\
16.07.96 & NGC~4501 & LS+CCD $520\times580$ & 33 min &
$2\arcsec\times2.5\arcmin$ & 6200--7000~\AA\ & $46^\circ$ &
1.3\arcsec \\
16.07.96 & NGC~4501 & LS+CCD $520\times580$ & 20 min &
$2\arcsec\times2.5\arcmin$ & 6200--7000~\AA & $92^\circ$ &
1.3\arcsec \\
5.05.97 & NGC~4501 & MPFS+CCD $1040\times1160$ & 60 min &
$10\arcsec\times21\arcsec$ & 4800--5400~\AA\ & $1^\circ$ &
2.5\arcsec \\
2.05.97 & NGC~4216 & MPFS+CCD $1040\times1160$ & 60 min &
$10\arcsec\times21\arcsec$ & 4800--5400~\AA\ & $48^\circ$ &
2.5\arcsec \\
2.05.97 & NGC~4216 & MPFS+CCD $1040\times1160$ & 40 min &
$10\arcsec\times21\arcsec$ & 6200--6900~\AA\ & $48^\circ$ &
2.5\arcsec \\
21.01.98 & NGC~4501 & MPFS+CCD $520\times580$ & 40 min &
$10\arcsec\times16\arcsec$ & 6200--6900~\AA\ & $ 2^\circ$ &
2.4\arcsec \\
\tableline
\end{tabular}
\end{flushleft}
\end{table}

In July 1996 we observed NGC~4501 with the Long-Slit Spectrograph
of the 6m telescope, under rather good seeing conditions. The slit,
with a length of $2\farcm5$ and a width of 2\arcsec, has been set
in three position angles: along the major axis, along the minor axis,
and between them, at $45^\circ$ from the major axis. We exposed a
red spectral range
containing strong emission lines H$\alpha$ and [NII]$\lambda6583$
and derived line-of-sight velocities of the ionized gas along
the slit by measuring barycenters of these emission lines.
The dispersion was 1.56 \AA\ per pixel, and the spectral resolution
varied along the slit between 3.2 and 3.9 \AA.
Test measurements of the sky emission
line [OI]$\lambda6300$ have shown that the accuracy of our line-of-sight
velocity measurements is 6-8 km/s. In the center where the emission
lines are especially strong we have used one-pixel spatial bin equal to
$0\farcs4$, in more distant regions  -- three-pixel binning.

The more detailed spectral investigation of the innermost parts
of NGC~4216 and NGC~4501 has been undertaken with the Multi-Pupil
Field Spectrograph (MPFS) of the 6m telescope (for description of
the early variant of the MPFS -- see \cite{afetal90}). This type
of spectrographs is still
rather untraditional; a detailed explanation of all the profits of
bidimensional spectroscopy can be found in the paper of Bacon et al.
(\cite{betal95}), and some details of our particular approach are given
in our work (\cite{silvb97}). Briefly, we obtain simultaneously
about one hundred spectra, each spectrum being from a square spatial
element of $1\farcs3 \times 1\farcs3$. The total element array covers
a rectangular area in the center of a galaxy.
The basic data reduction steps -- bias subtraction, flatfielding,
cosmic ray hit removing, extraction
of one-dimensional spectra, transformation into wavelength scale,
construction of surface brightness maps and velocity fields --
are performed using the software developed
in the Special Astrophysical Observatory (\cite{vlas}). Observing
NGC~4216 and NGC~4501, we used two different setups. In May 1996
and in January 1998 a CCD of $520 \times 580$ was in work; we exposed
an area of $8 \times 12$ elements ($10\arcsec \times 16\arcsec$) with
a dispersion of 1.6--1.7~\AA\ per pixel (spectral resolution being
3--5~\AA). In May 1997 a CCD of $1040 \times 1160$ was operated, and
we exposed an area of $8 \times 16$ elements
($10\arcsec \times 21\arcsec$) with a dispersion of 0.9~\AA\ per pixel
and a spectral resolution of 1.5~\AA. As a rule, we observe two spectral
ranges: green, 4700--5400 \AA, containing strong absorption lines
H$\beta$, MgIb, FeI$\lambda5270$ and 5335, and red, 6200--6900 \AA,
containing strong emission lines, H$\alpha$ and [NII]$\lambda6583$.
The green spectral range is used to derive radial profiles of the
absorption-line indices Mgb, Fe5270, Fe5335, and H$\beta$ calculated
in the well-known Lick system (\cite{woretal}); to keep nearly
constant signal-to-noise ratio along the radius, we have integrated the
spectra in concentric circular rings with a width equal to one element
($1\farcs3$) which are centered onto a galactic nucleus (for highly
flattened systems,
such as NGC~4216, this procedure results in smoothing details
of the radial profiles). Also, we use the green spectral range to
construct velocity fields of the stellar components;
for this purpose we cross-correlate every galactic elementary spectrum
with a spectrum of a bright star, typically K0--K3 III, obtained
with the same instrumental setup. The emission lines in the red
spectral range are measured to construct velocity
fields of the ionized gas. We estimate the typical accuracy of
the absorption-line indices to be 0.1~\AA\ and an intrinsic velocity
accuracy $\sim 20$ km/s.

To study in detail the morphology of the central parts of NGC~4216
and 4501, we have used a collection of digital images of 113 galaxies
presented by Frei et al. (\cite{frei}). NGC~4216 and 4501,
among the others,
were observed at the 1.5m telescope of Palomar Observatory on May 4,
1991. Three images, through the $g$, $r$, and $i$ Thuan-Gunn filters,
were obtained with the CCD TI of $800 \times 800$. The exposure times
were taken equal to 60 sec, but the nuclei of both galaxies
appeared to be saturated on the $r$- and $i$-images. The scale was
$1\farcs19$ per pixel, quite appropriate under moderate seeing
conditions which accompanied the observations: Frei et al. (\cite{frei})
report seeing $FWHM$ equal to $2\farcs2$--$2\farcs6$ for NGC~4501 and
$2\farcs6$--$3\farcs4$ for NGC~4216. Besides these moderate-resolution
images, we have used an image of NGC~4501, obtained with the HST WFPC2
in the frame of the programme of Stiavelli "Core properties of the
bulges of spiral galaxies", on May 3, 1997. The galaxy was exposed
through the broad-band filter $F606W$, the total exposure time
was 600 sec (400+200). The scale on the frame PC1 which contains
the nuclear part of the galaxy is $0\farcs045$ per pixel, the
spatial resolution estimated by a $FWHM$ of foreground stars
is $0\farcs1$, and under such sampling and resolution the central
$35\arcsec \times 35\arcsec$ of the galaxy has been exposed.
The morphology of the images was analysed with the programme of
one of us (V.V.V.) FITELL.

\section{Morphology of the Central Regions of NGC~4216 and NGC~4501}

\subsection{Ionized gas}

Fig.~1 presents MPFS maps of the central parts of NGC~4216 and 4501
in the continuum and in the emission line [NII]$\lambda$6583.
It is clearly seen that in the center of NGC~4216 the emission-line
morphology matches the continuum morphology, and the ionized gas
distribution is similar to the distribution of stars.
We agree with Keel (\cite{keel}) who characterized
the emission distribution in the center of NGC~4216 as a gaseous
disk. Emission morphology in the center of NGC~4501 is less obvious,
but we state that it is not point-like as Keel (\cite{keel}) reported.
We see an extended circumnuclear gaseous structure, but indeed the
emission isophote shape in the center of NGC~4501 is more roundish
than that of the continuum isophotes. Moreover, it lacks an axial
symmetry: the northern part of this structure is fainter than
the southern part. In other words, the observed morphology of the
emission-line distribution in the
center of NGC~4501 does not allow us to determine the
spatial structure of the gaseous subsystem; it differs from
the stellar distribution, although the kinematics of the ionized gas
and stars in the center of NGC~4501 are quite similar (see the
next Section).

\subsection{Stars}

Figs.~2--3 present some results of the isophote analysis using the
central parts of the broad-band
images of both galaxies. Here, together with our
results, we plot K-band isophote characteristics from the work of
Rauscher (\cite{rauscher}) -- it is important for such dusty regions
as the centers of NGC~4216 and 4501. In NGC~4216 there is no isophote
major axis twist: a slight increase of $PA$ toward the center
according to the data of Frei et al. (\cite{frei}) is not confirmed
by the results of Rauscher (\cite{rauscher}) in the radius range of
1.5\arcsec--3\arcsec. But other important features are evident: a local
maximum of ellipticity at $R\approx 11\arcsec$ and highly positive
(particularly in the bluer filters) values of the fourth cosine Fourier
coefficient $a_4$ at $R\lesssim 11\arcsec$. The bulge subtracted
photometric residual maps clearly show a highly inclined nuclear
stellar disk with a radius of $\sim$10\arcsec--11\arcsec\ which
lies exactly in the
plane of the main galactic disk. Rubin et al. (\cite{rubin97})
mentioned a weak bar in NGC~4216. But if it exists, it must lie
outside the radius range under consideration; in the central part of
NGC~4216 no bar signatures are detected: edge-on bars look usually
boxy, and in NGC~4216 within $R \approx 30\arcsec$ there
are no areas of negative $a_4$. The edge of the nuclear stellar
disk at $R\approx 10\arcsec$ can also be seen on the $H$-band surface
brightness profile given for NGC~4216 in the work of Jungwiert et al.
(\cite{jung97}). Jungwiert et al. (\cite{jung97}) looked for bars,
either single or double, in disk galaxies; for this purpose they have
made near-IR 2D photometry of 72 nearby galaxies and have analysed
morphological
characteristics of isophotes after deprojection of the images onto
the galactic planes. NGC~4216 was not deprojected because of its
large inclination, we could consider only the observed profiles.
However NGC~4501 was done, and Jungwiert et al. (\cite{jung97})
concluded that in this galaxy there are no "bar or other triaxiality",
and slight oscillations of morphological characteristics
near $R\approx 18\arcsec$ are caused by spiral arms. Carollo et al.
(\cite{car98}) do not analyse their HST image of NGC~4501
because of multiple spiral arms penetrating into the very center
of the galaxy. We have attempted to do this; together with the image
characteristics derived from the Frei's et al. (\cite{frei}) data and the
results of Rauscher (\cite{rauscher}), the morphological characteristics
of the isophotes of the HST WFPC2 image allow to make more detailed
conclusions about the structure of the center of NGC~4501. The apparent
ellipticity reaches a maximum value corresponding to a thin disk
inclined by $60^\circ$, namely, $\epsilon$=0.5, already at
$R\approx 7\arcsec$, in the bulge-dominated region. Besides, the
fourth cosine Fourier coefficient $a_4$ is always positive within
$R\approx 9\arcsec$ according to the HST data and within
$R\approx 13\arcsec$ according to the data of Frei et al. (\cite{frei})
and becomes negative outside this zone. Oscillations of $PA$
plus-minus $6^\circ$ from the line of nodes, seen even in the K-band,
are more than probable within the zone of strong $a_4$ variations.
So we detect morphological signatures of a circumnuclear stellar disk
inside a radius of 10\arcsec\ and a slightly
triaxial bulge dominating outside. As for nuclear "dust"
spirals, they are clearly seen in the residual HST image of NGC~4501
obtained by subtracting a model image consisting of pure elliptical
isophotes (Fig.~4). However, together with the "dark" spirals,
"luminous" spirals are seen too. Perhaps, the SW, more luminous, arm
causes the asymmetry visible in the emission-line image of the center
of NGC~4501 (Fig.~1).

\section{Kinematics of the Central Regions of NGC~4216 and NGC~4501}

\subsection{Stellar components}

Rotation of the stellar components in NGC~4216 and NGC~4501 has not yet
been studied. We have obtained two-dimensional velocity fields
of stars for the central regions of $10\arcsec \times 21\arcsec$ of
both galaxies by using the green spectral range data from MPFS.
They are shown in Fig.~5.

First of all, two-dimensional velocity fields allow us to check if
the rotation of stars is axisymmetric and if the plane of rotation
coincides with the main plane of the galactic disk. When we have
a disk which density
(brightness) distribution and rotation are axisymmetric,
the azimuthal dependence of the central line-of-sight velocity gradients
must follow a cosine law within the solid-body rotation radius: \\

\noindent
$dv_r/dr = \omega$ sin $i$ cos $(PA - PA_0)$, \\

\noindent
where $\omega$ is the deprojected angular rotation velocity, $i$ is the
inclination of the rotation plane, and $PA_0$ is the orientation of the
line of nodes. Since the axisymmetric disk is round, the line of nodes
is aligned with the isophotes. So if the maximum of the
cosine-like azimuthal dependence of the central line-of-sight velocity
gradients stays at the photometric major axis orientation, it strongly
hints for an axisymmetric rotation.

Fig.~6 presents the azimuthal dependence of the central line-of-sight
velocity gradients for NGC~4216; it is cosine-like.
Table~2 contains parameters of the
best-fitting cosinusoid calculated by the least-square method. One
can see that the orientation of the kinematical major axis is
$201^\circ$. The orientation of the main major axis of the galaxy
was known to be $199^\circ$ (RC3). We have obtained from the
morphological analysis in the previous Section that the outermost
isophotes have $PA_0=201^\circ$ and that the isophote orientation
in the center is practically the same. So we conclude that the stars
in the center of NGC~4216 exhibit axisymmetric rotation. Taking into
account that the galaxy is seen at an inclination of $i=78^\circ$
(\cite{guthrie}), we estimate the mean central angular rotation velocity
as 28 km/s/arcsec (or 350 km/s/kpc). The best circular-rotation model
applied to the complete two-dimensional velocity field has given
consistent values: $PA_0=200^\circ$ and $i=74^\circ$.

\begin{table}
\caption[ ] {Stellar rotation parameters for the center of NGC~4216}
\begin{flushleft}
\begin{tabular}{cll}
\tableline
Radius range, & Amplitude, & $PA_0$,\\
\arcsec &  $\omega \,\sin i$, km/s/arcsec& degrees\\
\tableline
1.5-2.4 &  27.4 & 201\\
\tableline
\end{tabular}
\end{flushleft}
\end{table}

\begin{table}
\caption[ ] {Stellar rotation parameters for the center of NGC~4501}
\begin{flushleft}
\begin{tabular}{cll}
\tableline
Radius range, & Amplitude, & $PA_0$,\\
\arcsec &  $\omega \, \sin i$, km/s/arcsec & degrees\\
\tableline
May 96 & & \\
1.5-2.3 &  29.6 & 138\\
2.7-4.2 &  24.7 & 128.5 \\
All 1.5-4.2 & 26.2 & 133.6\\
May 97 & & \\
$\leq 2$  &  28.3 & 126 \\
2.6-3.7&  27.0 & 138 \\
3.9-5.5&  19.5 & 139 \\
All 1-3.7  &  27.3   & 134.1 \\
\tableline
\end{tabular}
\end{flushleft}
\end{table}

Fig.~7 presents the azimuthal dependencies of the central line-of-sight
velocity gradients for NGC~4501 according to the observational data
from two runs, in May 1996 and May 1997. They are also cosine-like.
Table~3 contains the parameters of the best-fitting cosinusoids. First
of all, one can see that beyond $R \approx 4\arcsec$ the cosine
amplitude drops; that means that $R \approx 4\arcsec$ is the border of
the solid-body rotation region. Within $R \approx 4\arcsec$ the mean
value of $PA_0$ is $134^\circ \pm 3^\circ$ which is not too different
from the line-of-nodes orientation of $140^\circ$ (RC3; our isophote
analysis gave us $141^\circ$). Therefore we classify the rotation
of the stellar component in the center of NGC~4501 as axisymmetric too.
It implies that the angular velocity is $31 \pm 1$ km/s/arcsec
(or 383 km/s/kpc) assuming $i \approx 60^\circ$.

Assuming that the rotation of stars in the centers of NGC~4216
and NGC~4501 is axisymmetric, we have calculated azimuthally averaged
rotation curves in the frame of the circular rotation model. They
are presented in Fig.~8. The parameters $i$ and $PA_0$ were
determined by fitting the model to the whole velocity fields. So
for NGC~4501, the value of $PA_0$ is slightly different from that
found by the cosinusoid analysis within the solid-body rotation area.
However, this value of $PA_0$ is even closer to the main line
of nodes. One can see that in Fig.~8 the region of solid-body rotation
in NGC~4216 extends to $R \approx 9\arcsec$, just
the same as the radius of the circumnuclear gaseous disk (Rubin et al.
\cite{rubin97}); so the central part of NGC~4216 is kinematically
decoupled from the more outer part of the galaxy. The radius of the
solid-body rotation region in NGC~4501
is $R \approx 4\arcsec$ (Fig.~8). The local maximum rotation velocity
of $\sim$ 130 km/s at $R\approx 4\arcsec$ is followed by a slight
velocity decrease. So we conclude that inside the radius of the
kinematically distinct core in NGC~4501, 4\arcsec\ or 320 pc, the mass
is approximately $10^9\,M_{\sun}$ -- a quite typical value
for the kinematically distinct cores in spiral galaxies (\cite{asz89}).

\subsection{Ionized gas}

The rotation of the ionized gas in NGC~4216 and NGC~4501 was studied
more than once. We can mention the works of Rubin et al.
(\cite{rubin89}) and Woods et al. (\cite{woods}). These studies are
fully based on long-slit spectra taken along the major axes;
besides, they were intended
to obtain large-scale, extended rotation velocity distributions, so
their spatial resolution was rather low: Rubin et al. (\cite{rubin89})
used a spatial binning of $1\farcs5$ per pixel, and Woods et al.
(\cite{woods}) observed under the seeing conditions of
$FWHM \approx 3\arcsec \div 5\arcsec$. Such an approach has allowed
Rubin with co-authors to
find a kinematically distinct circumnuclear gaseous disk with a
radius of 10\arcsec\ in NGC~4216 (Rubin et al. \cite{rubin97}), but
a much more compact kinematically distinct core in NGC~4501 was not
noticed.

For the first time we present two-dimensional velocity fields of
the ionized gas in the centers of NGC~4216 and 4501 (Fig.~9). If
the gas velocity field for NGC~4501 looks quite regular and reminds
us of the stellar component velocity field (Fig.~5b),
the gas velocity field in the center of NGC~4216 is somewhat
asymmetric. Taking into account
that the galaxy is highly inclined and that we see the circumnuclear
gaseous disk almost edge-on, we now wish to further study
the gas rotation using the major-axis velocity measurements. For this
purpose we have simulated "long-slit cross-section" by superposing
a slit-like mask $1\farcs3 \times 20\arcsec$ on the two-dimensional
velocity field. The results -- one-dimensional velocity profile
along the major axis -- are presented in Fig.~10 together with
the previous observations of Woods et al. (\cite{woods}) and of
Rubin et al. (\cite{rubin97}) (the systemic velocity of 130 km/s
is subtracted from the measured line-of-sight velocities).
Except for two points at
$R=6\arcsec-8\arcsec$ to the south-west from the center, the
agreement is good. But in general, this
line-of-sight velocity profile does not look regular. For the NE part
the agreement is better, but Rubin et al. (\cite{rubin97})
noted emission-line splitting at some radii. To the south-west
from the nucleus the splitting was not noticed, but all three
measurements diverge at $R > 2\arcsec$, implying that there may be
some spatial sharp variations of the line-of-sight velocity pattern.
Kuijken and Merrifield (\cite{km95}) have simulated kinematical
signatures of bars from edge-on to end-on; they have
concluded that LOSVDs both of gas and stars (under the assumption
of smooth spatial gas and star density distributions) must
demonstrate gaps at some velocities giving a characteristic
"figure-of-eight" picture. We have checked emission-line profiles
(for the strongest emission line [NII]$\lambda$6583) along
the major axis of NGC~4216. With our spectral resolution, 1.5~\AA\
or 75 km/s, we clearly see LOSVD asymmetry with a long tail
toward systemic velocity which is characteristic of the
axisymmetric plane rotation seen almost edge-on; but we do not
see any gaps in the profiles. So we conclude that there are
no kinematical signatures of a bar within $R=10\arcsec$
in NGC~4216 -- this conclusion is consistent with the results
of morphological analysis presented in the previous Section.
The irregular shape of the major-axis velocity profile may be a
consequence of a patchy dust distribution; some north-south asymmetry
is also seen in the emission-line brightness map (Fig.~1a).
An additional argument in favor of axisymmetric gas rotation in
the center of NGC~4216 is the full agreement between the angular
gas rotation velocity --
28 km/s/arcsec -- reported by Rubin et al. (\cite{rubin97}) (which
is in fact line-of-sight velocity gradient along the major axis)
and the stellar angular rotation velocity measured in the previous
subsection; in the presence of a bar the stars (collisionless dynamic
system) and the gas (dissipational dynamic system) should have
different distributions and kinematics.

\begin{table}
\caption[ ] {Gaseous rotation parameters for the center of NGC~4501}
\begin{flushleft}
\begin{tabular}{clll}
\tableline
Radius range, & Emission line & Amplitude, & $PA_0$,\\
\arcsec & & $\omega \, \sin i$, km/s/arcsec & degrees\\
\tableline
May 96 & & & \\
2.0-2.4 &  [NII]6583 & 28.9  & 147\\
2.7-4.2 &  [NII]6583 & 20.4 & 126 \\
Jan 98 & & & \\
2.0-2.8 & [NII]6583 &  24.2 & 134 \\
2.0-2.8 & $H_{\alpha}$ & 25.7 & 146 \\
2.0-2.8 & [SII]6717 & 29.5 & 134 \\
3.3-4.7& [NII]6583 & 16.4 &132 \\
All & & &\\
2.0-2.8  &    &  26.8   & 141 \\
\tableline
\end{tabular}
\end{flushleft}
\end{table}

Fig.~11 presents the azimuthal dependencies of central line-of-sight
velocity gradients for the ionized gas of NGC~4501 in two different
radius ranges and from measurements of various emission lines.
The parameters of fitted cosinusoids are given in Table~4.
First of all, one can see that the amplitude of the cosinusoids drops
noticeably beyond $R \approx 3\arcsec$: it is the border of the
solid-body rotating gaseous disk. The data of May 1996 and January
1998 are in agreement, and to determine parameters of the nuclear
gas rotation, we involve measurements of [NII]$\lambda$6583 in 1996
and of [NII]$\lambda$6583, H$\alpha$, and [SII]$\lambda$6717 in 1998.
The orientation of the dynamical major axis for the ring of
$R=2\arcsec$--$3\arcsec$ is $PA_0$=$141^\circ \pm 4^\circ$; it
suggests that the rotation is axisymmetric, because the orientation of the
main line of nodes is $PA_0=140^\circ$ (RC3) or $141^\circ$ (our
analysis). The averaged
angular rotation velocity is $31 \pm 1$ km/s/" (or $388 \pm 17$
km/s/kpc) if we assume the inclination to be $60^\circ$. We
see that in the center of NGC~4501 both stars and ionized gas
rotate axisymmetrically, in the equatorial plane of the galaxy,
and with the same angular speed.

Fig.~12 presents three long-slit cross-sections of NGC~4501: these
profiles supplement major-axis cross-sections published earlier
by Rubin et al. (\cite{rubin89}) and by Woods et al. (\cite{woods}).
One can note that along the minor axis there is no velocity gradient
near the center of NGC~4501; this is one more evidence for axisymmetric
rotation of the circumnuclear ionized gas in NGC~4501. Fig.~13
shows the rotation curve for the ionized gas in the center of
NGC~4501, both the azimuthally averaged one constructed from the
two-dimensional MPFS data in the frame of an axisymmetric-rotation
model and the one derived from the linear long-slit cross-sections.
Up to $R \approx 3\arcsec$
the MPFS and long-slit rotation curves agree perfectly despite very
different spatial resolution (seeing conditions) of two data sets,
$1\farcs3$ for the long-slit data and $2\farcs4$ for the MPFS data. But
beyond $R \approx 4\arcsec$ two branches of the long-slit major-axis
cross-sections diverge extremely. By modelling the one-dimensional
major-axis cross-section from the two-dimensional MPFS velocity field,
we have obtained the same asymmetry. The azimuthally averaged MPFS
rotation curve lies between these two branches, just as
one could expect. The gaseous disk thus seems strongly asymmetric:
perhaps, to the south-east
from the center at $R \geq 3\arcsec$ we see the circumnuclear disk,
and to the north-west -- something else (dust screen?) that hides
the NW part of the circumnuclear disk.
If so, we must take into account only the SE branch of the major-axis
cross-section, then the local maximum of the gas rotation velocity
shifts to $R \approx 7\arcsec$, and the gaseous circumnuclear disk
in NGC~4501 appears to be more extended that we thought before.

\section{Chemically Distinct Nuclei in NGC~4216 and 4501}

Figs.~14 and 15 present the radial profiles of the absorption-line
indices for NGC~4216 and NGC~4501, respectively. These profiles give
us the evidence for a chemical distinctness of the nuclei in both
galaxies: one can see a drop of metal-line equivalent widths when
passing from the nuclei to the bulges. In the bulges which we
take safely at $R \geq 4\arcsec$ (let us remind that seeing quality was
$2\farcs5$) there are no noticeable gradients of metal-line indices,
so we average the values in the radius range of 4\arcsec--9\arcsec.
Now we can make quantitative estimates of differences in stellar
populations properties by using models for old stellar populations.
In NGC~4216 the mean metal-line indices in the bulge are
$<\mbox{Mgb}>=4.20 \pm 0.05$ \AA\ and $<\mbox{Fe}>=2.72 \pm 0.06$
\AA\ ($<\mbox{Fe}>\equiv$ (Fe5270+Fe5335)/2); so the differences
"nucleus--minus--bulge" are $\Delta \mbox{Mgb}=0.54 \pm 0.15$ \AA\
and $\Delta <\mbox{Fe}>=0.74 \pm 0.16$ \AA.
If we assume equal ages of the nuclear and bulge stellar populations
and if magnesium-to-iron ratio is solar, then by using the models
of Worthey (\cite{worth94}) we obtain $\Delta$[Fe/H]=$0.26 \pm 0.07$
from $\Delta$Mgb and $\Delta$[Fe/H]=$0.41 \pm 0.09$ from
$\Delta <\mbox{Fe}>$. The difference between these estimates suggests
a possible difference between magnesium-to-iron ratios in the nucleus
and in the bulge. Before we make analogous calculations for NGC~4501,
let us note that in the nucleus of NGC~4501 ($R < 3\arcsec$) a quite
measurable emission line [NI]$\lambda$5199 is seen in the spectra.
As this line contaminates the red continuum passband of the Mgb index,
its presence results in a Mgb overestimate. The effect has been
discussed and quantified by Goudfrooij \& Emsellem (\cite{ge96}); we
apply their recommended correction, 1.13EW([NI])=0.45~\AA, to our
nuclear Mgb measurement. After this correction, the bulge
characteristics being
$<\mbox{Mgb}>=4.26 \pm 0.08$ \AA\ and $<\mbox{Fe5270}>=3.08 \pm 0.13$
(according to the measurements of May 1997), we obtain the index
differences $\Delta \mbox{Mgb}=0.56 \pm 0.18$ \AA\ and
$\Delta \mbox{Fe5270}=0.62 \pm 0.23$ \AA\ and, under the same
assumptions as for NGC~4216, the metallicity difference $0.27 \pm 0.09$
from $\Delta$Mgb and $0.38 \pm 0.14$ from $\Delta$Fe5270.
Although the emission spectra of the nuclei of NGC~4216 and
NGC~4501 are quite different, the properties of the stellar populations
both in the nuclei and in the bulges appear to be very similar.

But two main assumptions, that of equal ages and that of solar
Mg-to-Fe ratio both in the nuclei and in the bulges, which allow to use
the models of Worthey (\cite{worth94}), are not necessarily valid.
Let us try to
clarify the situation in NGC~4216 and NGC~4501. For this purpose we can
use "index-index" diagrams; but to refer to these diagrams, we need
absolute values of the indices in the Lick system. Our spectral
resolution in May 1997 was 1.5~\AA\ -- much better than that of the
Lick system (8~\AA). So we may expect that our index measurements
in May 1997 are overestimated. Indeed, the differences between
the measurements of NGC~4501 in May 1997 and in May 1996 have the right
sense (Fig.~15). To check this effect, we observed bright stars
(4 stars --
in May 1997, resolution 1.5\AA, 9 stars -- later, resolution 4\AA) from
the Lick library (\cite{woretal}). We have found that with the higher
spectral resolution we overestimate the values of Mgb by 0.50~\AA\
and the values of Fe5270 by 0.44~\AA; with our standard resolution,
which was used in May 1996, the measured index values for the stars
coincide with tabular ones within the errors of individual
observations in the Lick project. Interestingly, just the same
differences, 0.4--0.5~\AA, can be seen between the May 1996 and the
May1997 measurements in Fig.~15, so we conclude that the
corrections cited above must only be applied to the measurements of
May 1997 to transform them into the Lick system.

Fig.~16 presents the diagrams (Fe5270, Mgb) for the central parts
of NGC~4216 and NGC~4501. One can see that these giant spiral galaxies
differ strongly from a bulk of giant ellipticals on this diagram:
if the latters lie to the right from the model sequences [Mg/Fe]=0
(\cite{worth92}), the measurements for the nuclei and bulges
of NGC~4216 and NGC~4501 lie close to them. The nucleus of NGC~4216
has obviously solar magnesium-to-iron ratio, the bulge may be slightly
magnesium-overabundant. In NGC~4501, after applying the correction
for the [NI] emission to the circumnuclear Mgb indices, both the
nucleus and the bulge seem to lie close to Worthey's
(\cite{worth94}) model sequences calculated for [Mg/Fe]=0.

As the assumption of solar magnesium-to-iron ratio appears to be valid
for the central parts of NGC~4216 and 4501, we can try to separate
metallicity and age effects on the diagrams (H$\beta$, Mgb),
(H$\beta$, [MgFe]) ([MgFe]$\equiv (\mbox{Mgb}<\mbox{Fe}>)^{1/2}$), or
(H$\beta$, $<\mbox{Fe}>$). Unfortunately the absorption line H$\beta$
in the center of NGC~4501 is strongly contaminated by emission. But
in the center of NGC~4216 the emission $H\beta$ is practically absent
 -- its equivalent width is less than 0.1~\AA, -- and we can
estimate an age of the stellar population in the nucleus and
its environment. Fig.~17 presents the comparison of our index data for
the center of NGC~4216 with the models of Worthey (\cite{worth94}) on
the diagrams (H$\beta$, Mgb) and (H$\beta$, [MgFe]) and with the
models of Tantalo et al. (\cite{tantalo})
for [Mg/Fe]=0 on the diagram (H$\beta$, $<\mbox{Fe}>$). Different
metal absorption-line indices and different models give a stable
result: the stars in the nucleus of NGC~4216 have a mean age of 8-12
billion years and a mean metallicity of $0.0 \div +0.1$, the stars
in the bulge
are older than 15--17 billion years and more metal-poor than the Sun.
So the chemically distinct nucleus in NGC~4216 is also younger
than the nearest bulge. As for quantitative self-consistent
estimates of the parameter differences, we can use recommendations
of Tantalo et al. (\cite{tantalo}): they give linear equations which
connect
index differences $\Delta \mbox{Mg}_2$, $\Delta <\mbox{Fe}>$, and
$\Delta \mbox{H} \beta$ with the stellar population parameter
differences $\Delta \log Z$, $\Delta$[Mg/Fe], and $\log t$. Assuming
$\Delta$Mgb=16.2$\Delta \mbox{Mg}_2$, we obtain for NGC~4216:
$\Delta$[Mg/Fe]=-0.12, $\Delta Z$=+0.48, and $\Delta \log t$=-0.39.
If we take the direct $\mbox{Mg}_2$ determinations from our spectra
for the nucleus and for the bulge of NGC~4216 according to the
prescriptions of Worthey et al. (1994), we obtain
$\Delta$[Mg/Fe]=-0.04, $\Delta Z$=+0.49, and $\Delta \log t$=-0.30,
quite in agreement with the former results.
So one can see that magnesium overabundance in the bulge is very
modest, not more than 0.1 dex, the nucleus is younger than the bulge
by a factor of 2--2.5, and, after taking into account the age
difference, the nucleus is more metal-rich than the bulge by a
factor of 3 instead of the factor of 2 obtained under the assumption
of equal ages.

\section{Conclusions}

By using bidimensional spectral data obtained at the 6m telescope
for the Virgo spirals NGC~4216 and NGC~4501, we have found chemically
distinct metal-rich nuclei in these galaxies. Under the assumption
of equal ages for the nuclear and bulge stellar populations, the
metallicity difference between the nuclei and their environments in
the galaxies is estimated as a factor of 2. But we have also found
an age difference between the nucleus and the bulge in NGC~4216:
age-metallicity disentangling results in an age estimate for the
nucleus of 8--12 billion years, the bulge being older by a factor
of 1.5--2; and the self-consistent metallicity difference estimate is
then a factor of 3. The solar magnesium-to-iron ratios in the
galactic nuclei suggest for a long duration of the secondary nuclear
star formation bursts which have produced the chemically distinct
stellar subsystems.

The detailed morphological and kinematical analyses
made for the stellar and gaseous structures in the centers of
NGC~4216 and 4501 have revealed the presence of the circumnuclear
stellar-gaseous disks which exhibit fast axisymmetric rotation and lie
in their respective equatorial planes of the main galactic disks.
The radius of the circumnuclear disk in NGC~4216 is 10\arcsec\
(800 pc), and this value agrees with the estimate of Rubin et al.
(\cite{rubin97}) made for the purely gaseous inner disk. The inner disk
radius in NGC~4501 can only be roughly estimated in the range
4\arcsec--7\arcsec\
due to the complex dust distribution in the central part of the galaxy.
Unlike the elliptical galaxies with chemically distinct cores
where area of magnesium-index variations coincides with the inner-disk
extensions (\cite{sb95}), the chemically distinct nuclei in NGC~4216
and 4501 seem to be unresolved although the inner disks have quite
measurable extensions. This work presents the second evidence of this
sort; the first was the case of M~31 (\cite{silbv98}), but it
looked more striking -- a circumnuclear 100 pc stellar
disk in this galaxy extends up to $0\farcm5$ from the center, and
the chemically distinct nucleus is much more compact, less than
4\arcsec. One may note that the inner disks contribute less
than 20\%\ of the total light in the bulge-dominated areas so they
cannot affect significantly the integrated spectra of the stellar
populations. But in the elliptical galaxies the situation is similar,
meanwhile the disks are detectable by their magnesium-enhanced
contributions. In any case, the unresolved chemically
distinct nuclei in spiral galaxies represent something
different from the inner disks, though there may exist some
genetic relation between these two substructures.

\acknowledgements
We thank the astronomers of the Special Astrophysical Observatory
Dr. S. N. Dodonov for supporting the observations at the 6m telescope
and Dr. V. H. Chavushyan for providing a possibility to get
some additional data in his observing time. Also we thank the
referee for the careful reading of the paper and very careful remarks.
The 6m telescope is operated under the financial support of
Science Department of Russia (registration number 01-43).
During the data analysis we have
used the Lyon-Meudon Extragalactic Database (LEDA) supplied by the
LEDA team at the CRAL-Observatoire de Lyon (France) and the NASA/IPAC
Extragalactic Database (NED) which is operated by the Jet Propulsion
Laboratory, California Institute of Technology, under contract with
the National Aeronautics and Space Administration. The work is partly
based
on observations made with the NASA/ESA Hubble Space Telescope, obtained
from the data archive at the Space Telescope Science Institute, which is
operated by the Association of Universities for Research in Astronomy,
Inc., under NASA contract NAS 5-2655.  We have used the software ADHOC
developped at the Marseille Observatory, France. The work
was supported by the grant of the Russian Foundation for Basic
Researches 98-02-16196 and by the Russian State Scientific-Technical
Program "Astronomy. Basic Space Researches" (the section "Astronomy").

\newpage

\figcaption{Surface brightness distributions of the continuum
at $\lambda$6500 (grey-scaled) and the emission [NII]$\lambda$6583
(isophotes) for the central parts of NGC~4216 ({\it a}) and
NGC~4501 ({\it b})}

\figcaption{Results of the isophote analysis for the broad-band
images of the central part of NGC~4216: {\it a} ---
major axis position angle radial variations, {\it b} ---
ellipticity radial variations, {\it c} --- boxiness radial
variations. Asymptotic parameter values at $R=120\arcsec -200\arcsec$
are: $PA_0=21\arcdeg$, $\epsilon \equiv 1-b/a =0.83$}

\figcaption{Results of the isophote analysis for the broad-band
images of the central part of NGC~4501: {\it a} ---
major axis position angle radial variations, {\it b} ---
ellipticity radial variations, {\it c} --- boxiness radial
variations. Asymptotic parameter values at $R=120\arcsec -160\arcsec$
are: $PA_0=141\arcdeg$, $\epsilon \equiv 1-b/a =0.54$}

\figcaption{The residual map (after pure elliptical isophote
subtraction) of the HST WFPC2 image of NGC~4501. The light ellipse
marks the region where model subtraction has been made;
the image is direct, $PA(top)=15\fdg3$,
the total image sizes are $36\arcsec \times 36\arcsec$. Luminous
(grey) and dark (white) spiral arms are clearly seen in the radius
range of $1\arcsec - 10\arcsec$}

\figcaption{Two-dimensional line-of-sight velocity maps for the
stellar component in the centers of NGC~4216 ({\it a}) and
NGC~4501 ({\it b}). The images are direct, the north
is pointed by an arrow, the photometric center is marked by a cross}

\figcaption{Azimuthal dependence of the central stellar line-of-sight
velocity gradient for NGC~4216}

\figcaption{Azimuthal dependence of the central stellar line-of-sight
velocity gradient for NGC~4501, in five radial bins}

\figcaption{Azimuthally-averaged rotation curves for the stars
in the centers of NGC~4216 and 4501 calculated over the observed
2D MPFS velocity fields in the frame of a
two-dimensional axisymmetric model;
the best-fit parameters for the orientation of the galaxies
are $PA_0=200\arcdeg$, $i=74\arcdeg$ for NGC~4216 ({\it a}) and
$PA_0=140\arcdeg$, $i=61\arcdeg$ for NGC~4501 ({\it b}). For
comparison the major-axis velocity profiles simulated by masking
our 2D velocity fields are also plotted as solid lines}

\figcaption{Two-dimensional line-of-sight velocity maps for the
ionized gas in the centers of NGC~4216 ({\it a}) and
NGC~4501 ({\it b}). The images are direct, the north
is pointed by an arrow, the photometric center is marked by a
cross}

\figcaption{The major-axis line-of-sight velocity profile for
the ionized gas ([NII]$\lambda$6583 emission line) in the
center of NGC~4216 simulated by masking our 2D velocity map.
For comparison the long-slit observations of Woods et al. (1990)
and of Rubin et al. (1997) are also plotted}

\figcaption{Azimuthal dependencies of the central line-of-sight
velocity gradient for the ionized gas in NGC~4501 in several radial
bins; solid lines show the best-fit cosinusoids with parameters
listed in Table~4}

\figcaption{Long-slit profiles of ionized-gas line-of-sight
velocities in $PA=137\arcdeg$ ({\it a}), $PA=46\arcdeg$ ({\it b}),
and $PA=92\arcdeg$ ({\it c}) for NGC~4501}

\figcaption{The ionized-gas rotation curve for NGC~4501.
The points with error bars are calculated from the 2D MPFS velocity
fields in the frame of a
two-dimensional axisymmetric model; the best fit
is reached for the orientation mentioned in the legend.
The long-slit major-axis measurements divided by
$\sin i$ are also plotted; a strong discrepancy
between the SE and NW parts of the major-axis cross-section is
observed at $R > 3\arcsec$}

\figcaption{Radial profiles of the absorption-line indices in
NGC~4216}

\figcaption{Radial profiles of the absorption-line indices in
NGC~4501}

\figcaption{The diagram (Fe5270, Mgb) for the central parts of
NGC~4216 ({\it a}) and NGC~4501 ({\it b}). The observational
points are taken along the radius with a step of $1\farcs3$ and
connected by dashed lines. The ages of the model stellar
populations are given in the legend in billion years}

\figcaption{The diagrams (H$\beta$, Mgb) ({\it a}),
(H$\beta$, [MgFe]) ({\it b}), and
(H$\beta$, $<\mbox{Fe}>$) ({\it c}) for the central part of
NGC~4216. The observational points are taken along the radius
with a step of $1\farcs3$ and connected by dashed lines. The ages
of the model stellar populations from the works of Worthey (1994)
({\it a,b}) and Tantalo et al. (1998) ({\it c}) are given in the
legend in billion years}

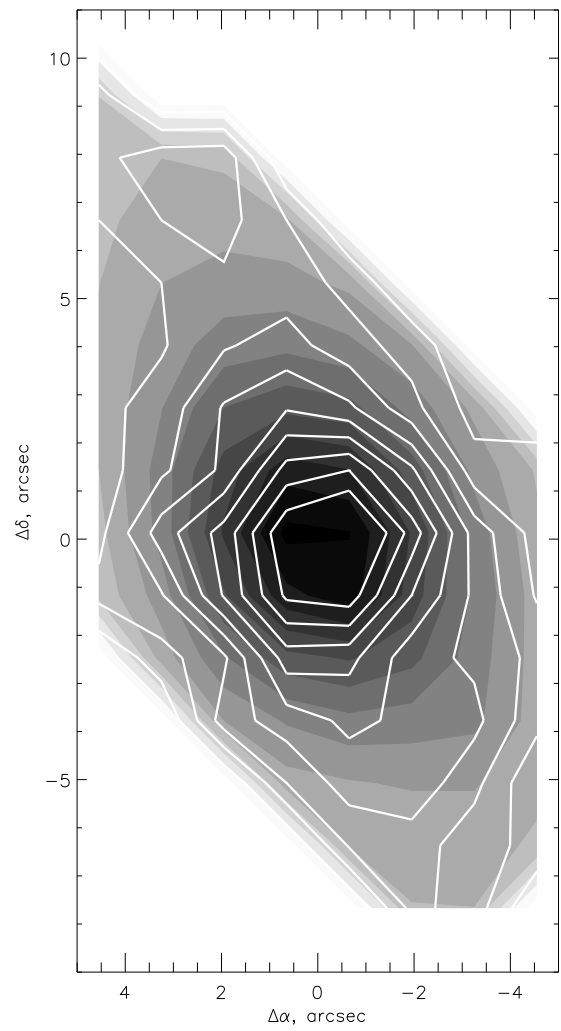
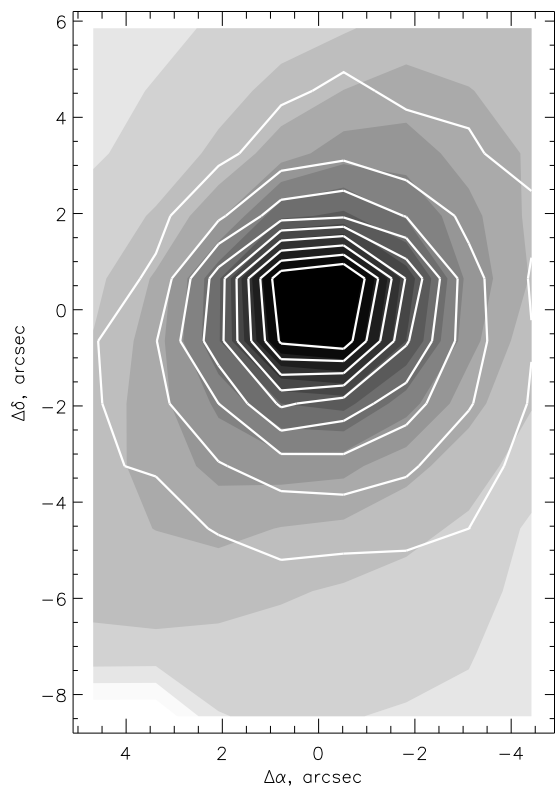
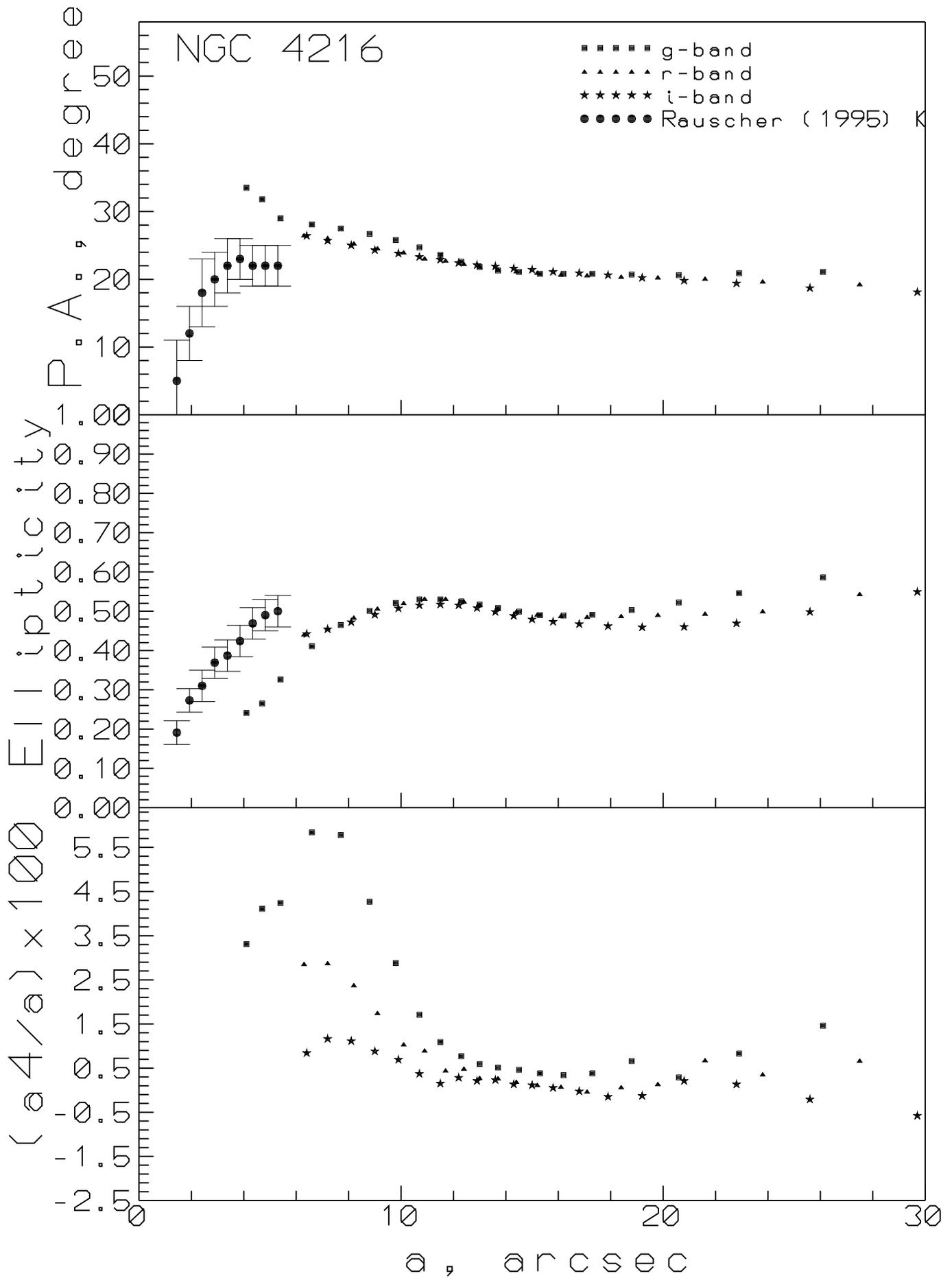
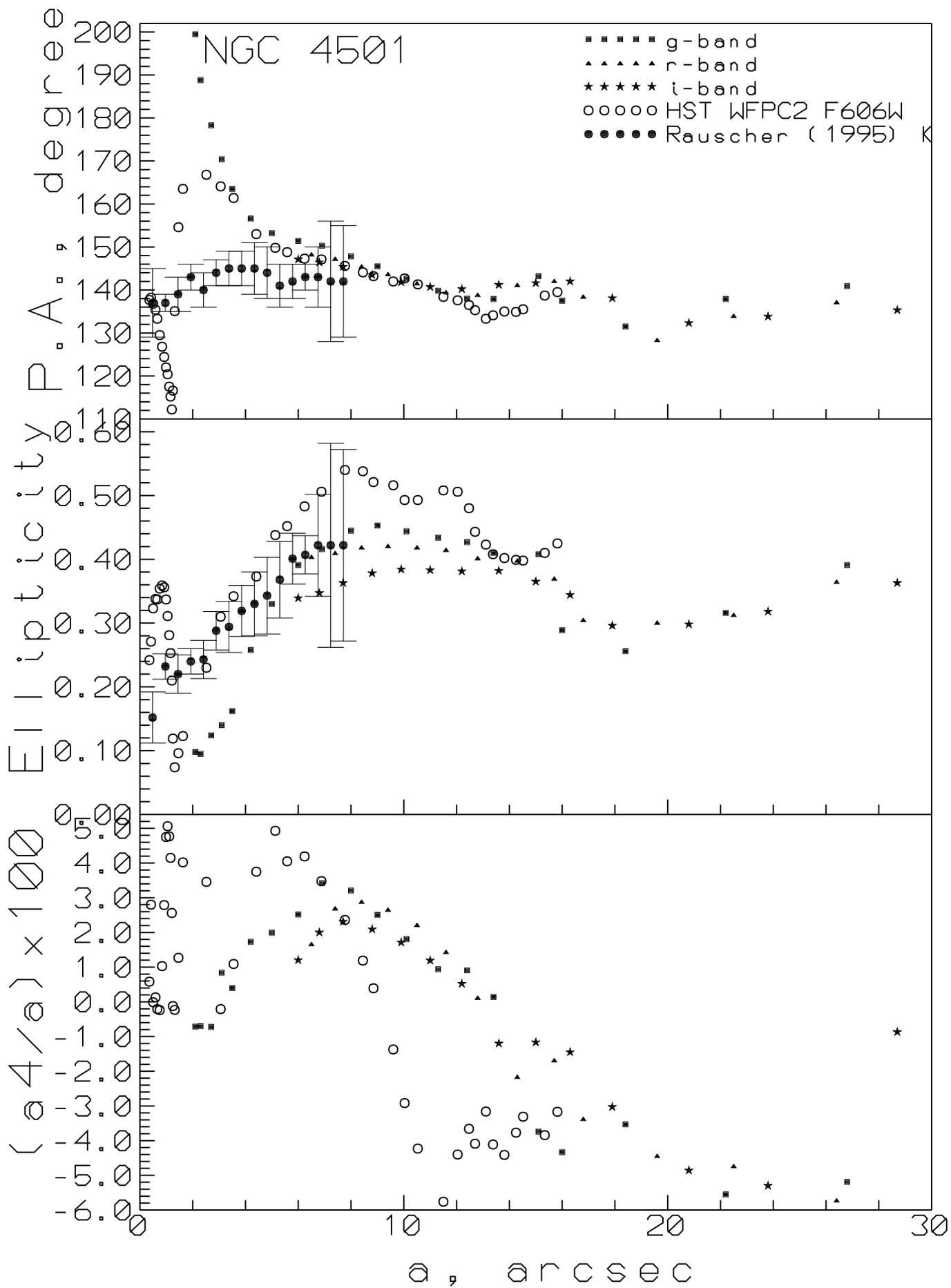
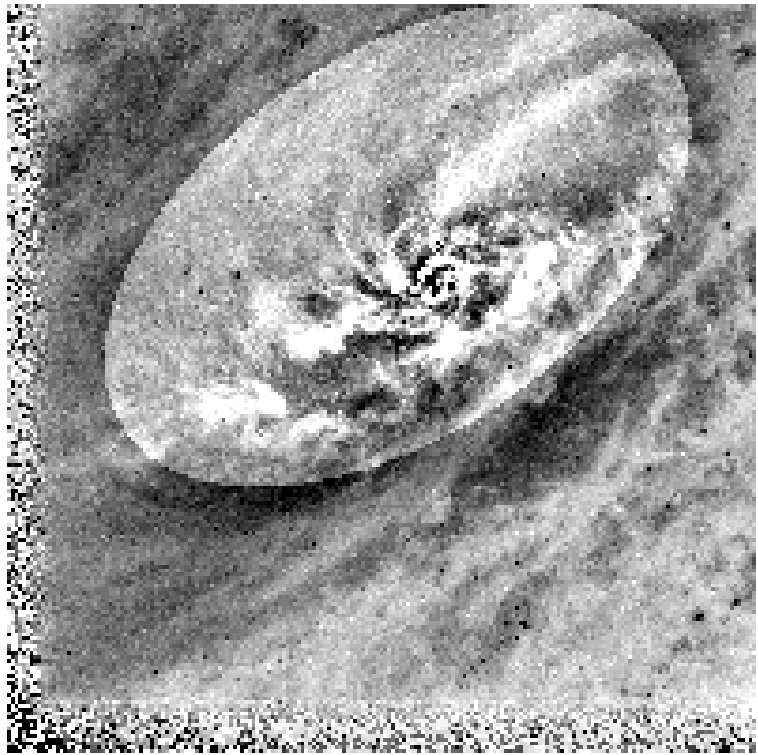
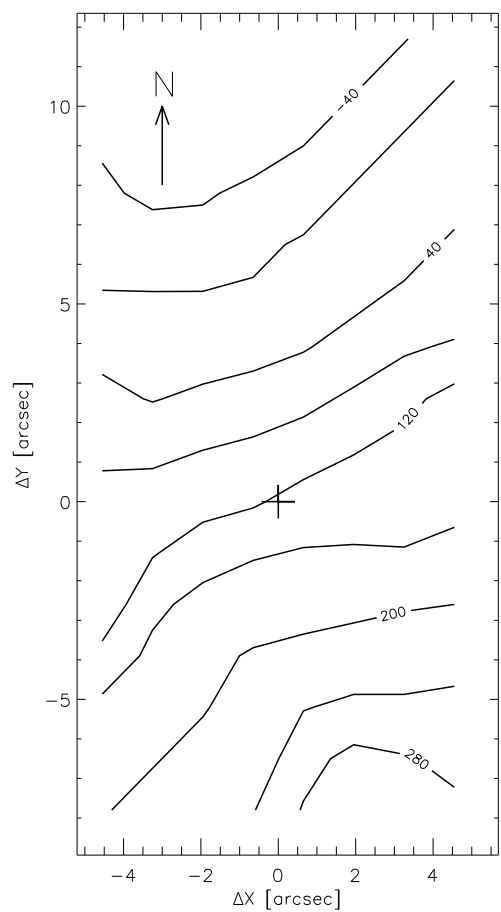
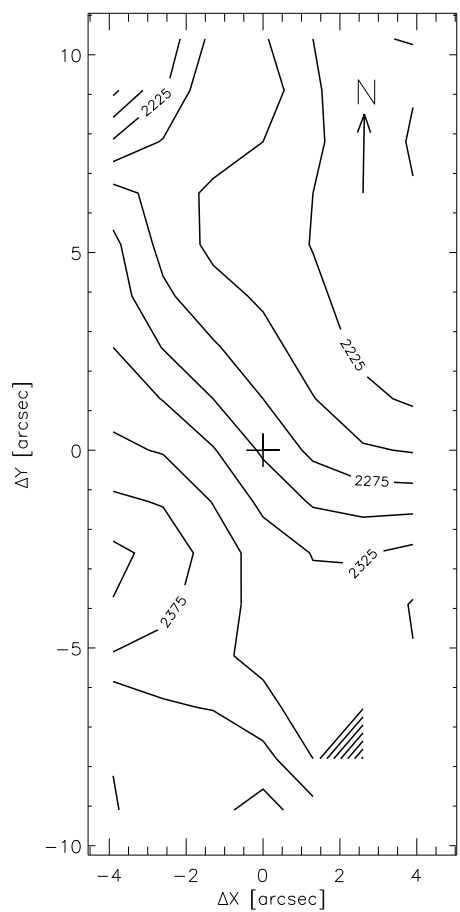
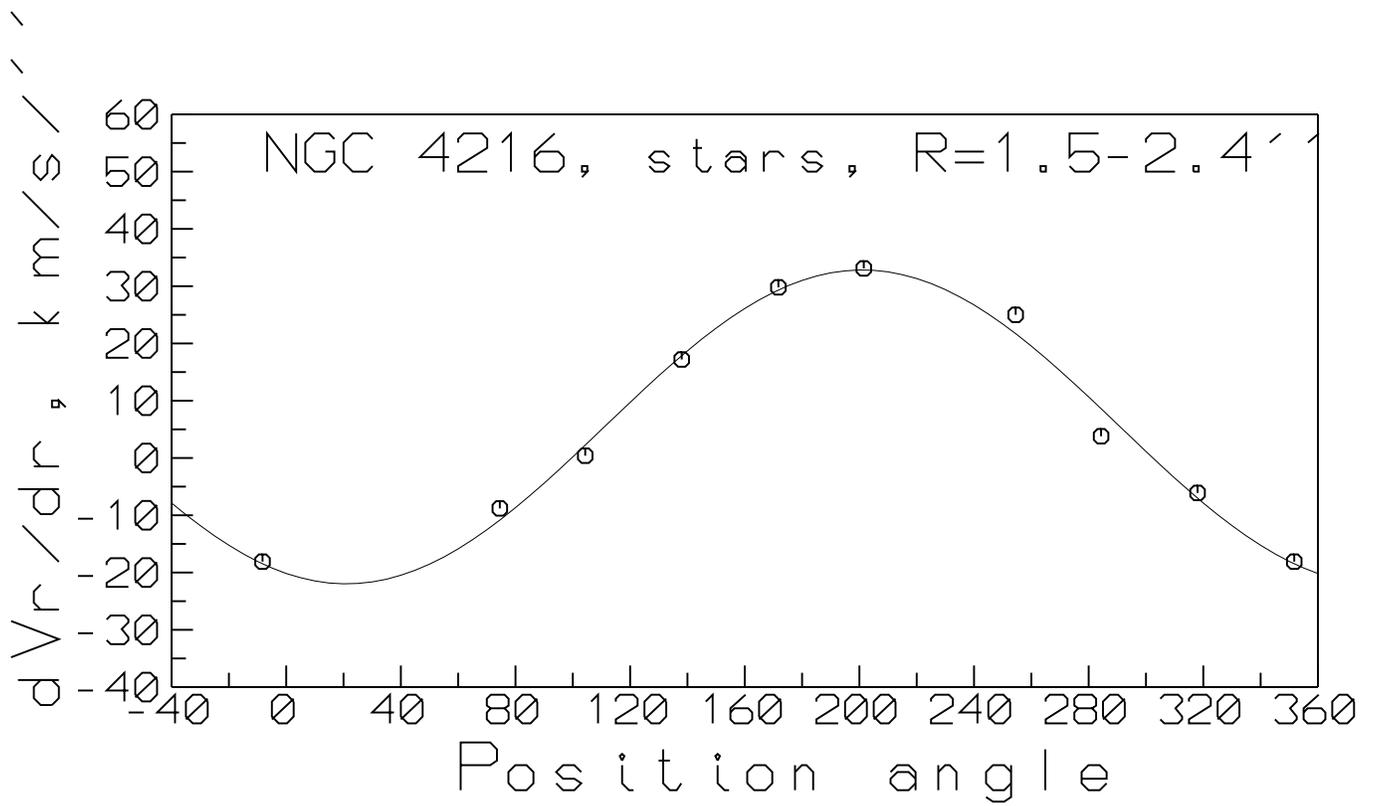
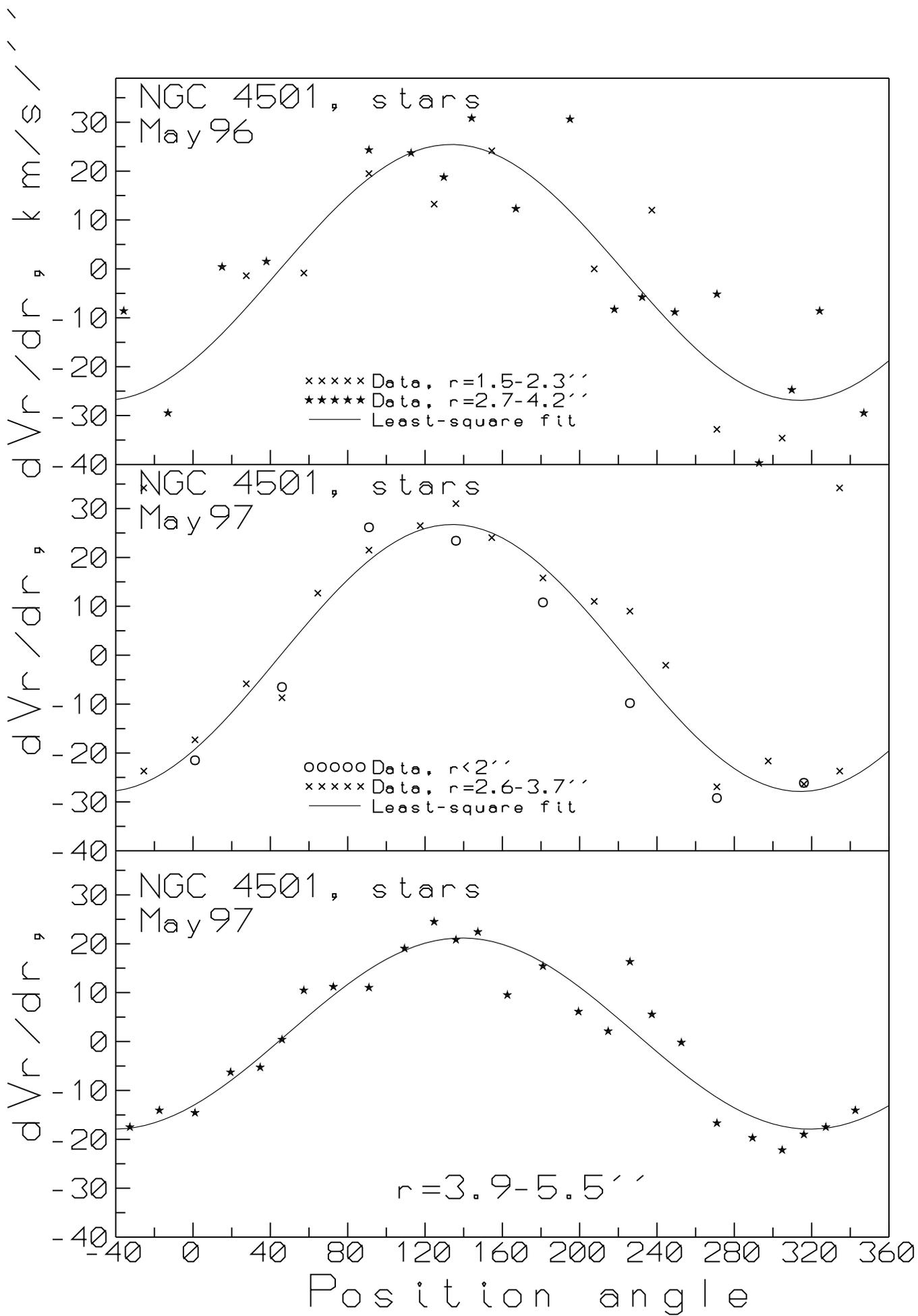
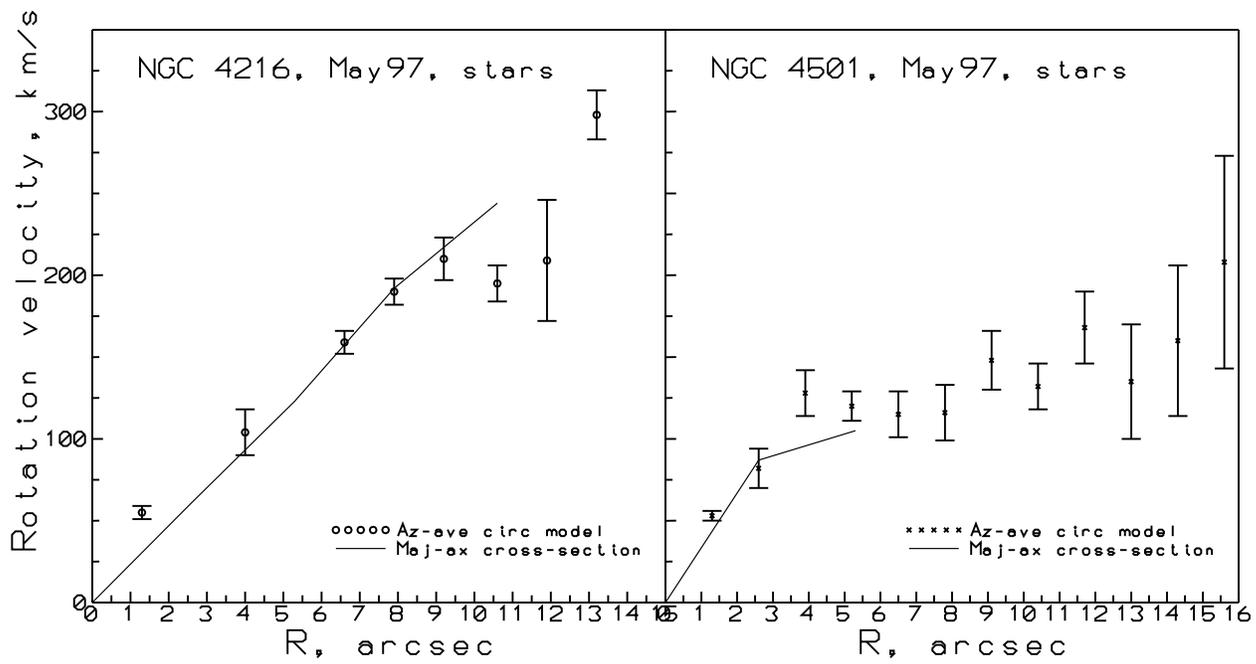
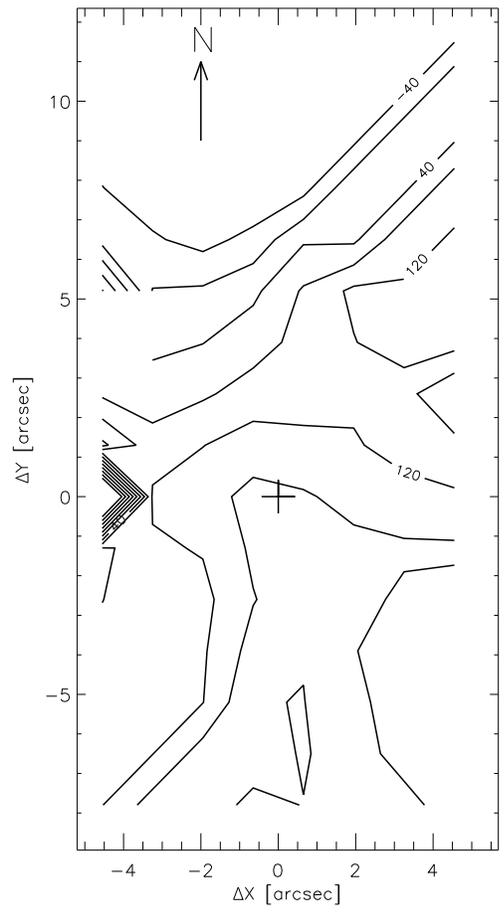
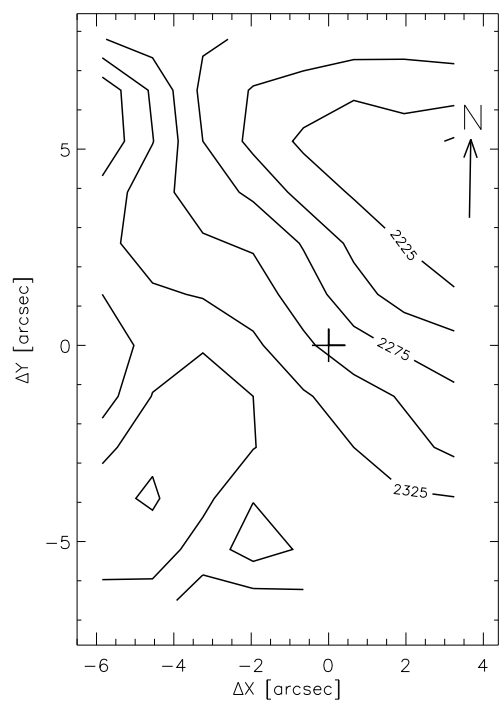
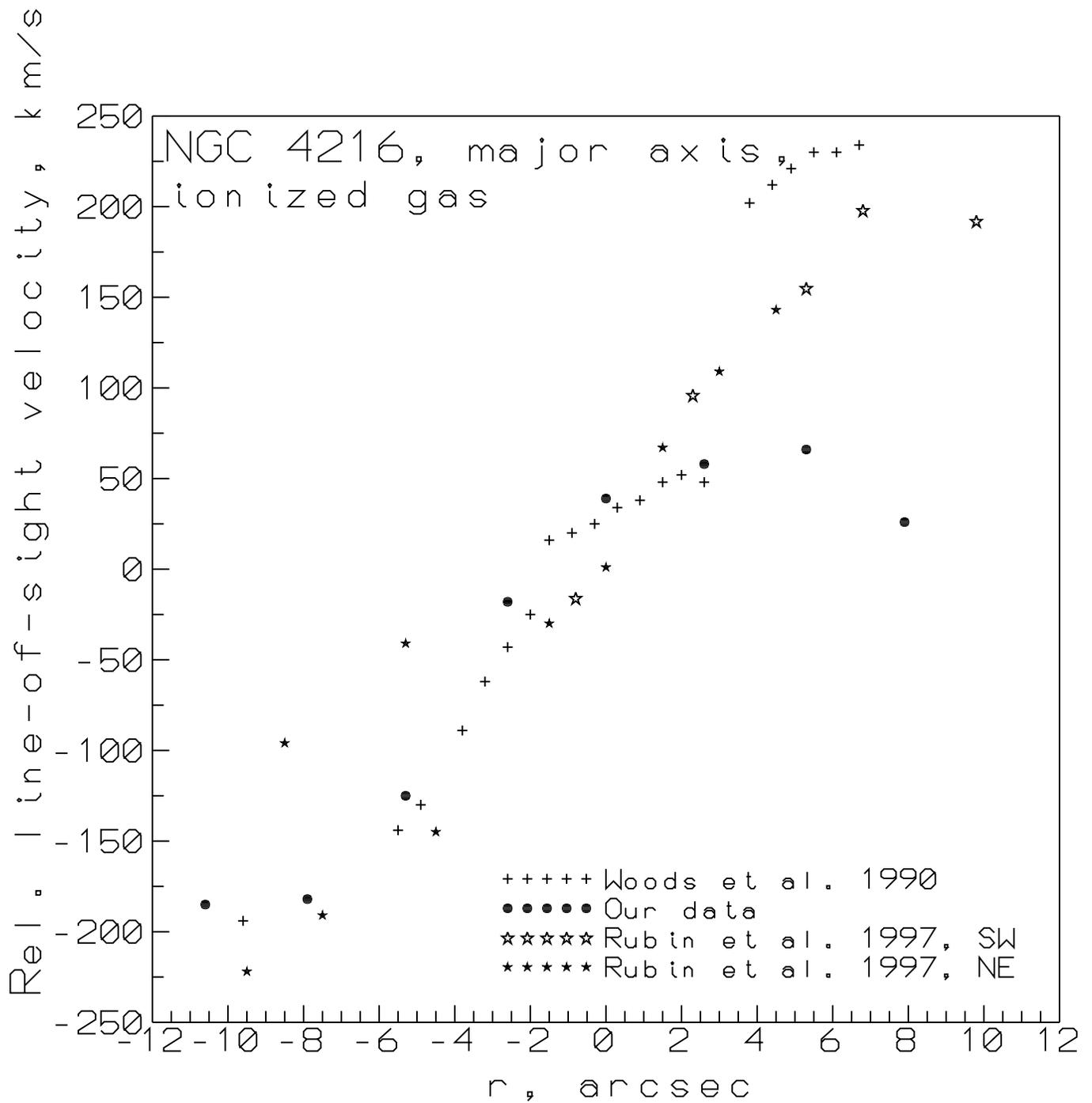
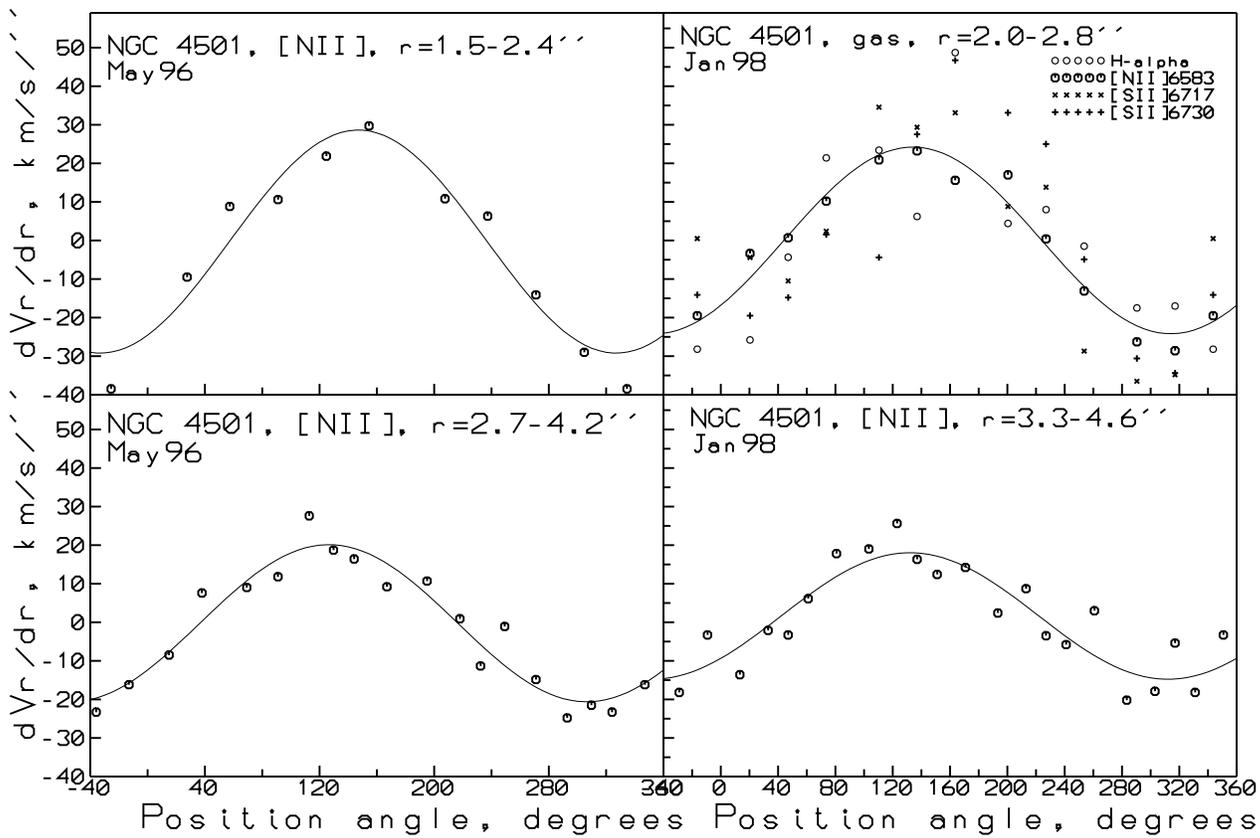
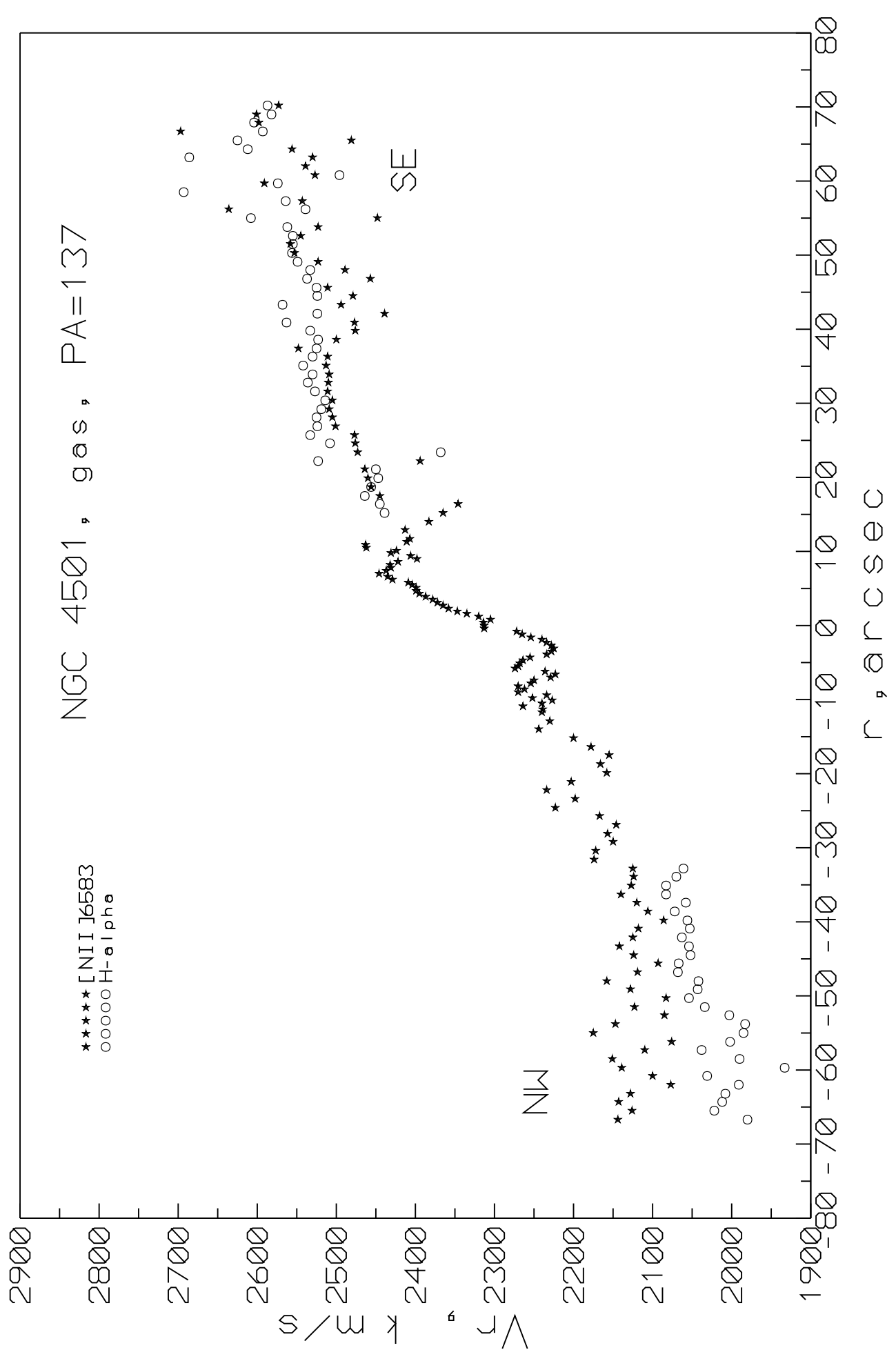
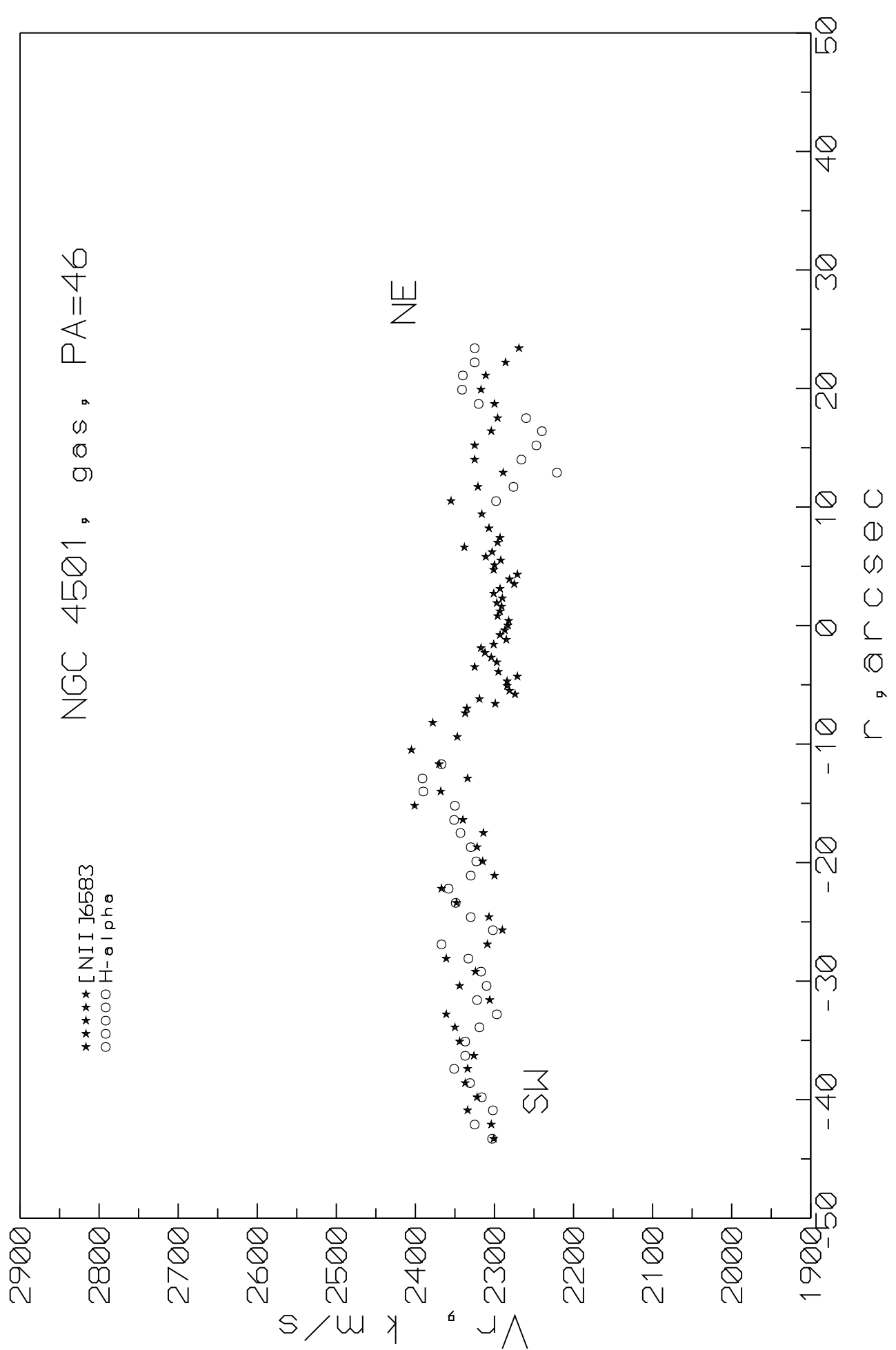
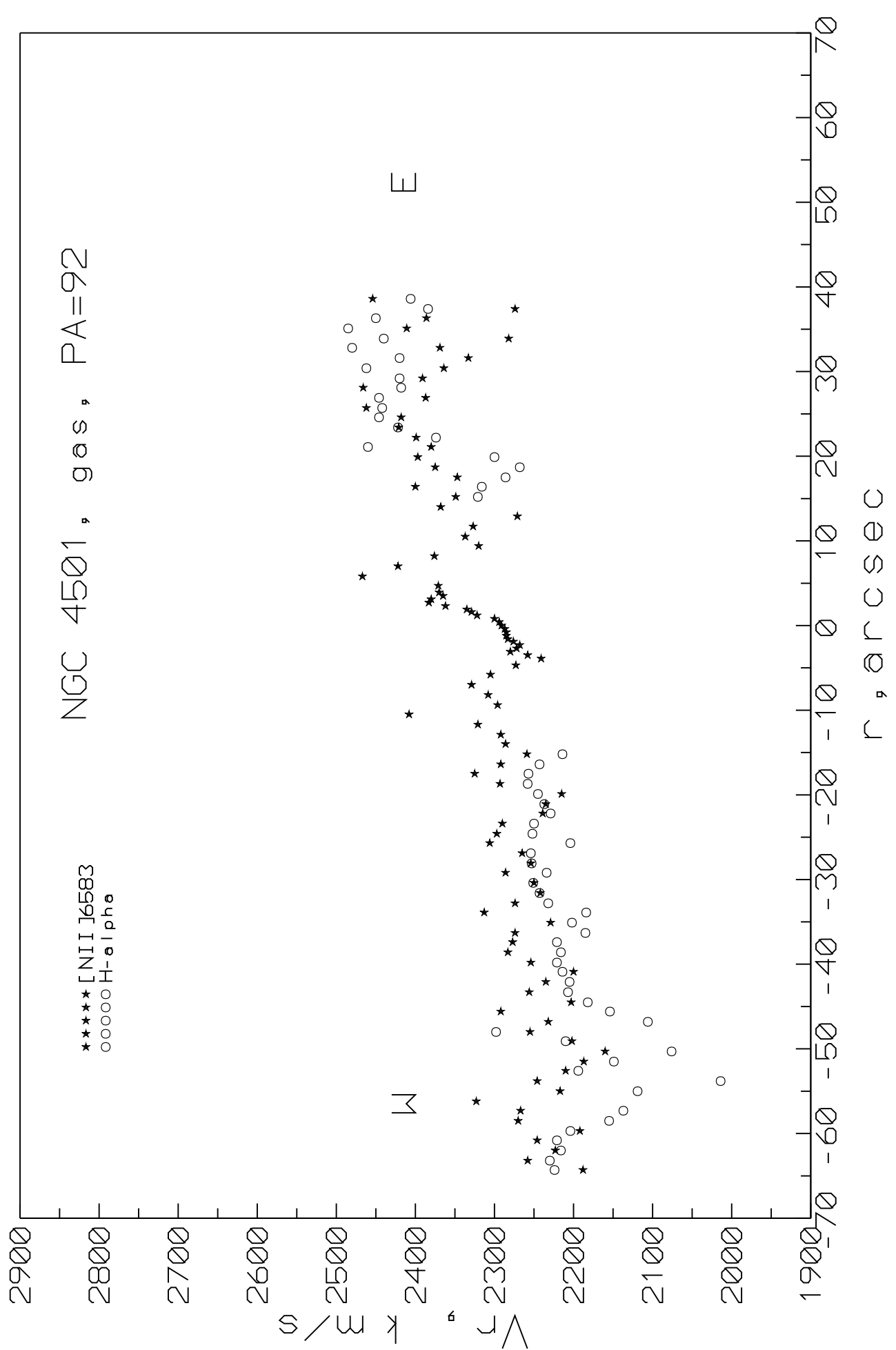
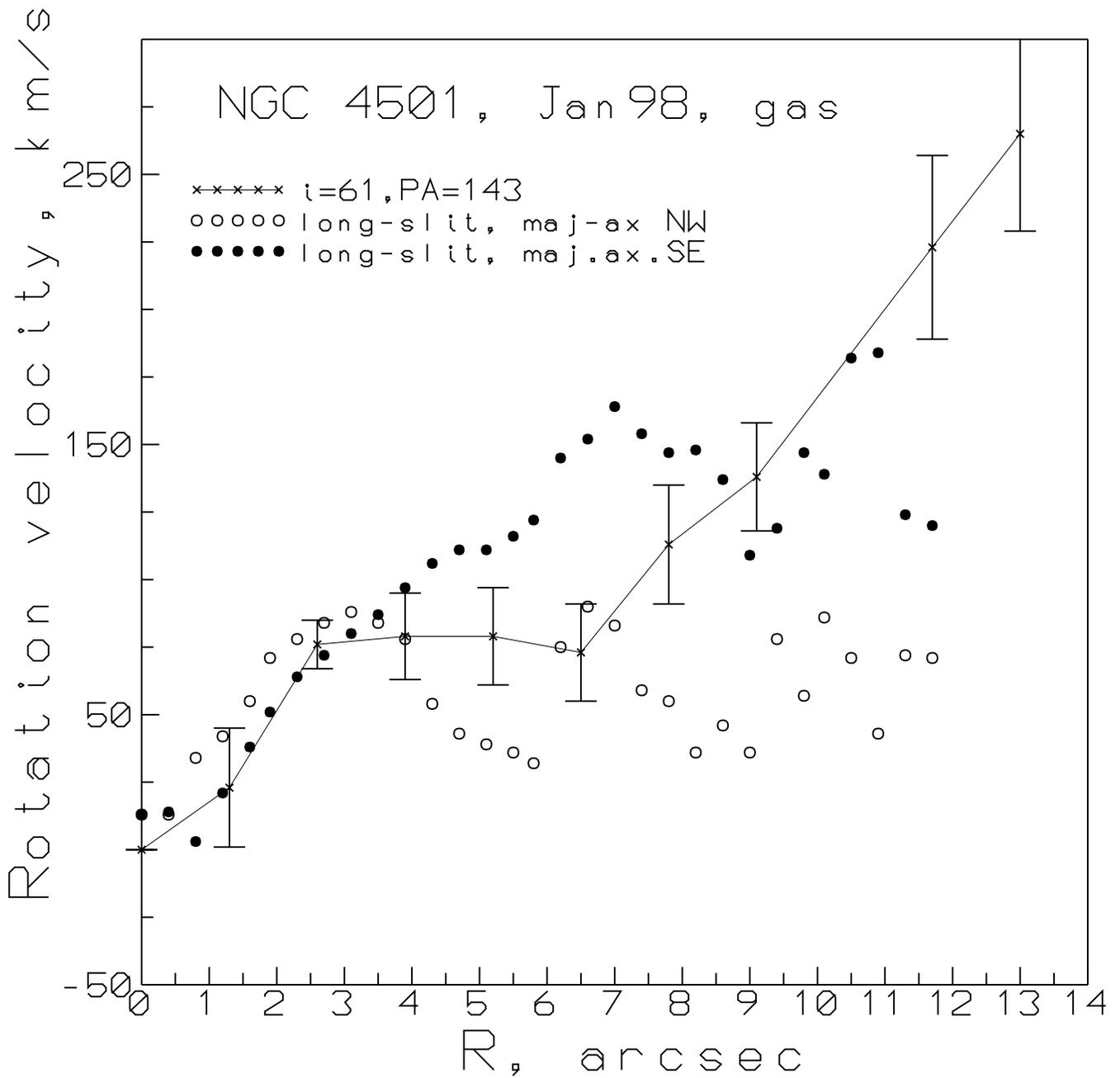
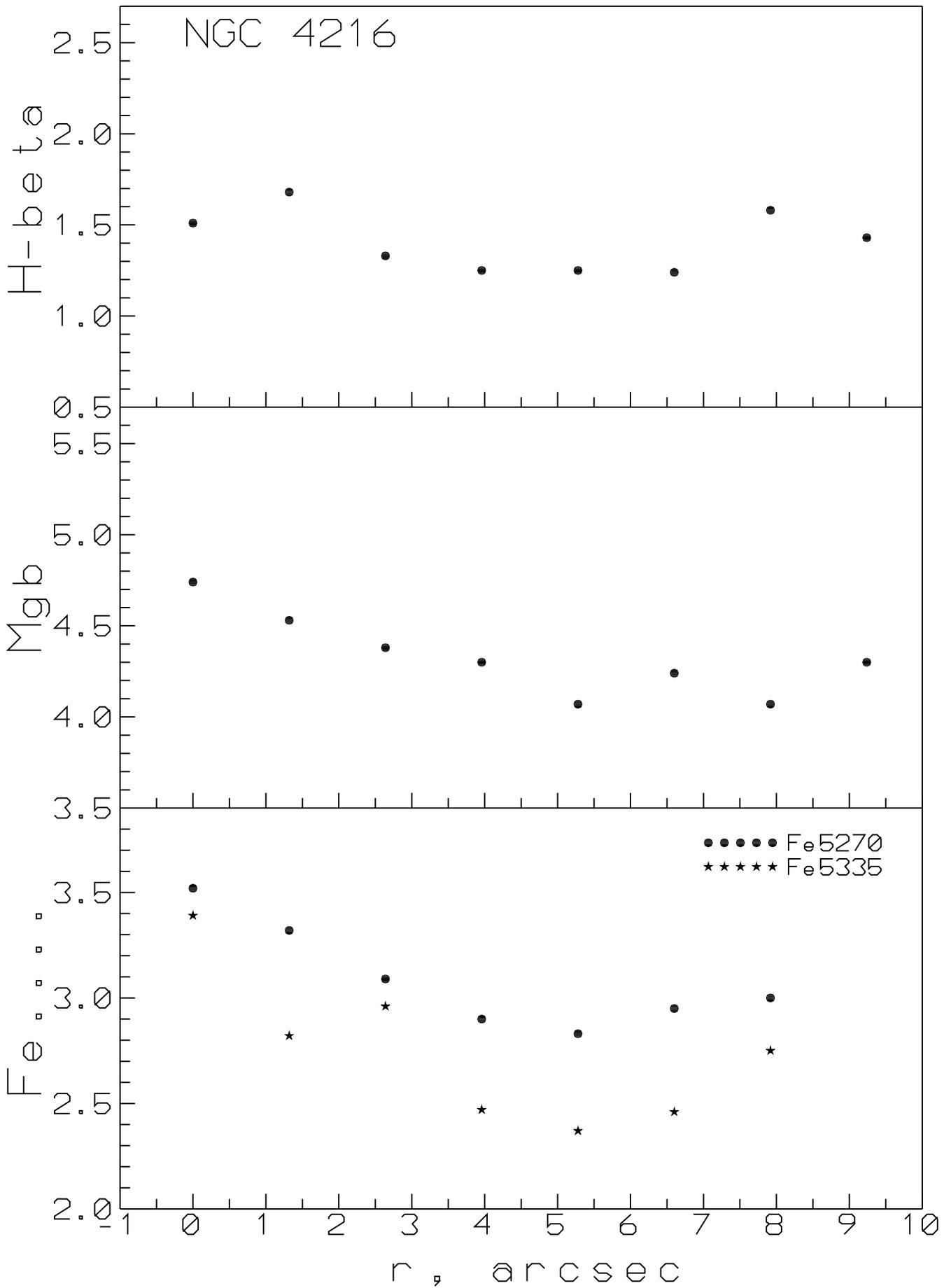
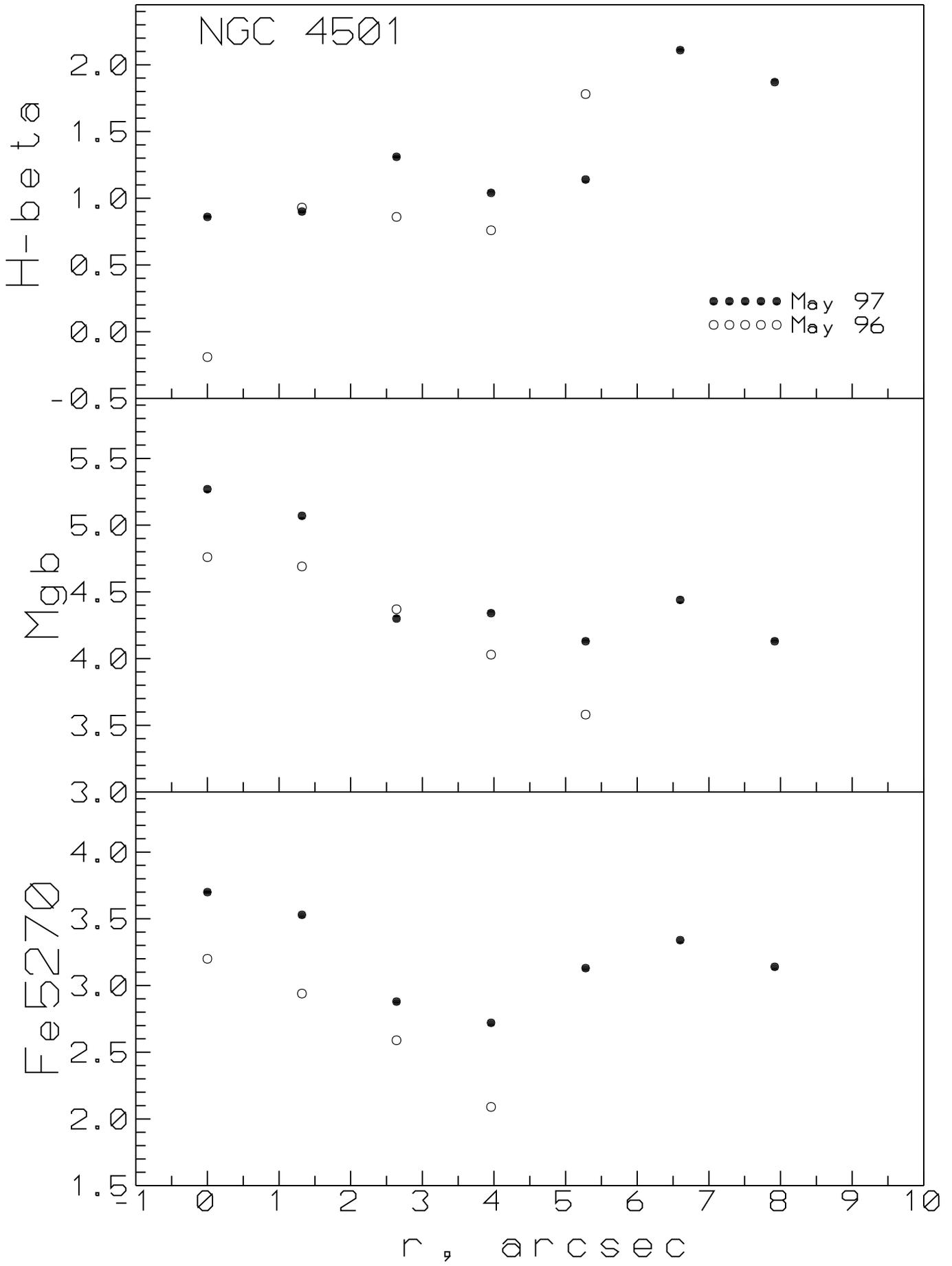
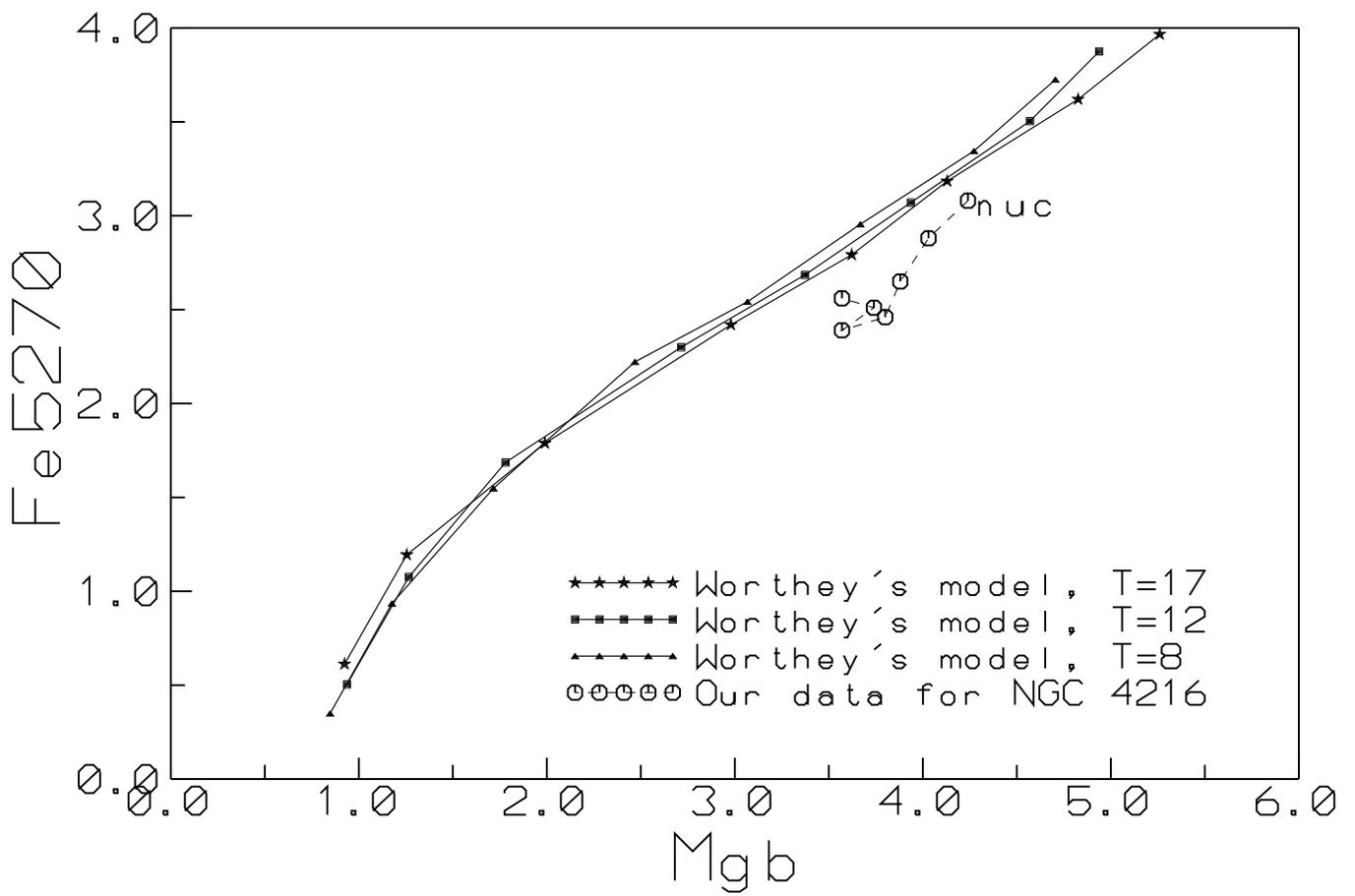
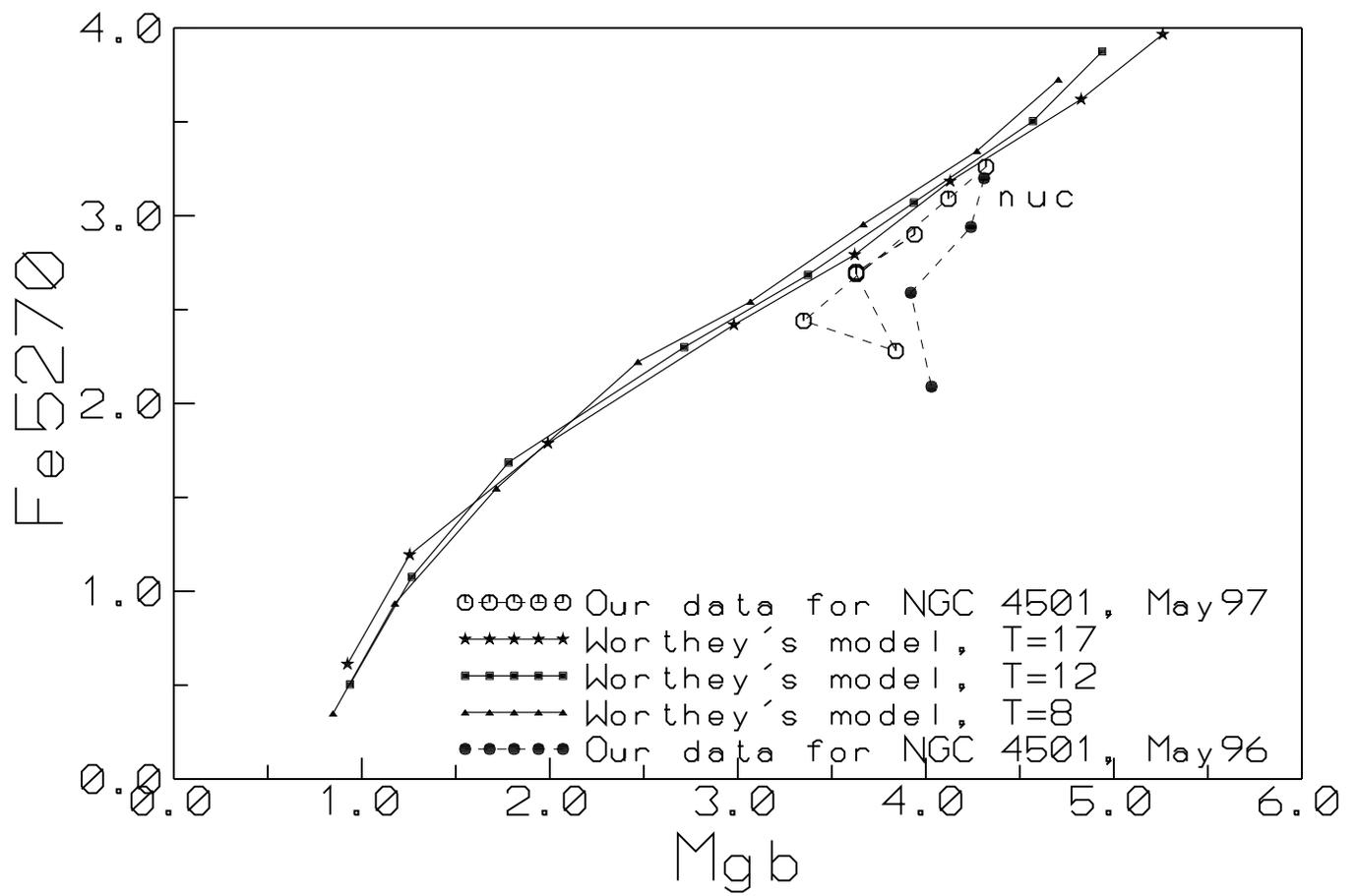
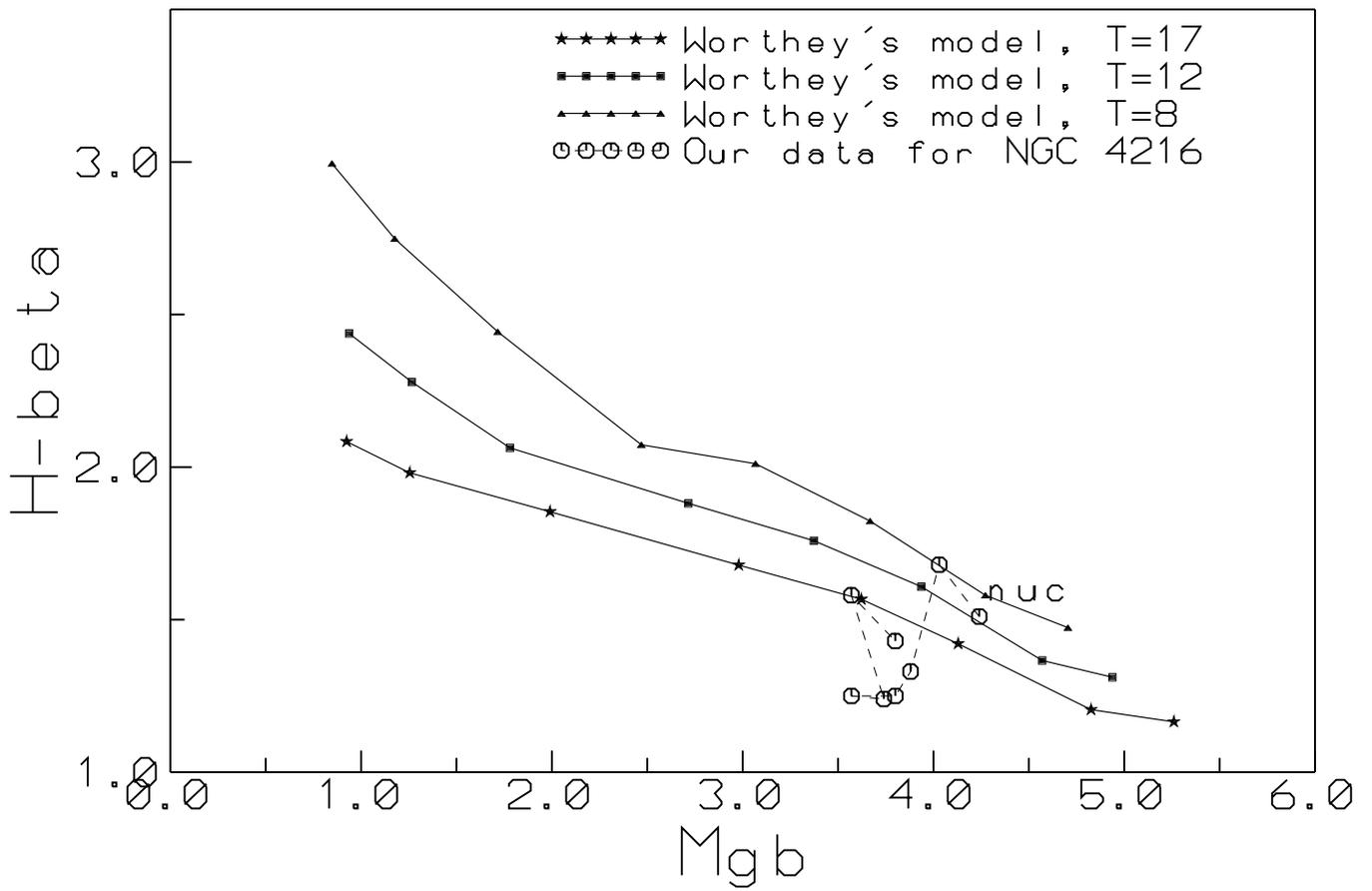
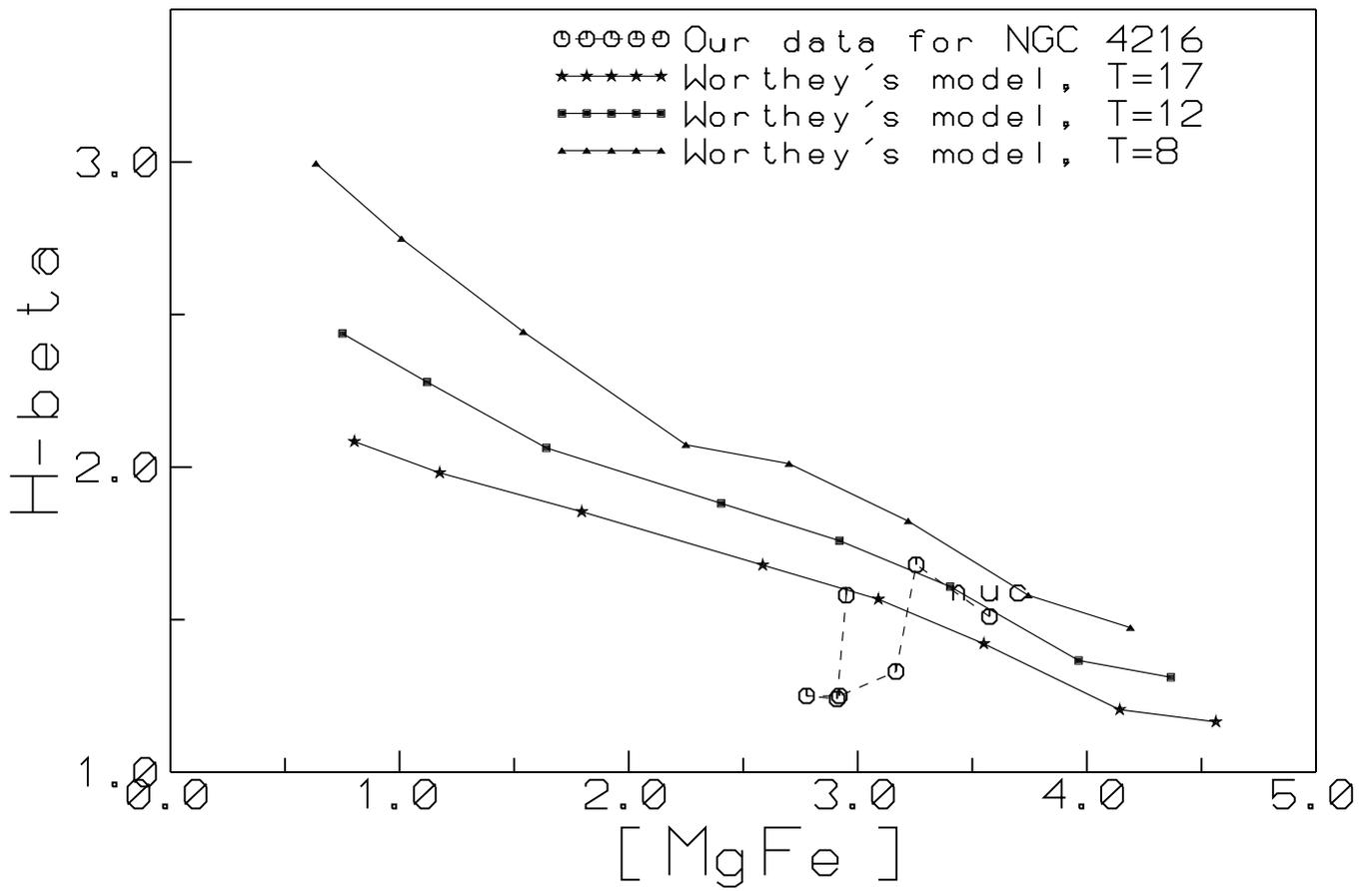
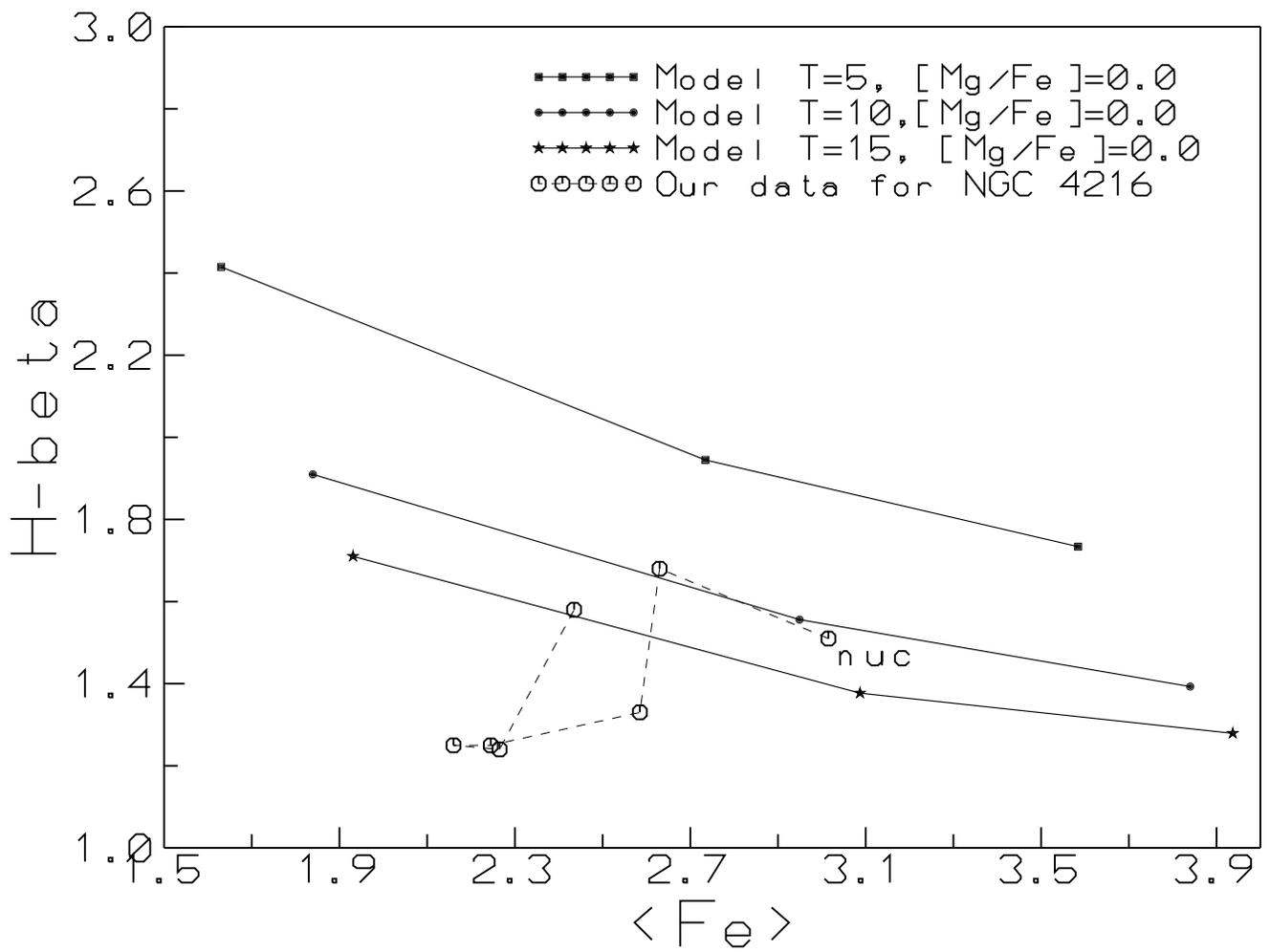
\end{document}